\documentclass[onecolumn]{emulateapj}
\def\IGR{IGR\,J17544-2619}
\def\xte{XTE\,J1739-302}
\def\igg{IGR\,J16479-4514}
\def\axj{AX\,J1841.0-0536}    
\def\igrjj{IGR\,J16465-4507} 
\def\igrj11{IGR\,J11215-5952}

\def\iigr{IGR\,J16418-4532}
\def\2s{2S\,0114+65}
\def\igrpatel{IGR\,J16358-4726}
\def\chan{{\em Chandra}}
\def\xmm{{\em XMM-Newton}}
\def\int{{\em INTEGRAL}}
\def\swift{{\em SWIFT}}
\shorttitle{Magnetars and SFXTs}

\begin{document}

\title{Are There Magnetars in High Mass X-ray Binaries? The Case of SuperGiant Fast X-Ray Transients}    

\author{E. Bozzo\altaffilmark{1,2}, 
        M. Falanga\altaffilmark{3},
        L. Stella\altaffilmark{1}
       }

\altaffiltext{1}{INAF - Osservatorio Astronomico di Roma, Via Frascati 33,
00044 Rome, Italy} 
\altaffiltext{2}{Dipartimento di Fisica - Universit\`a di Roma ``Tor
  Vergata'', via della Ricerca Scientifica 1, 00133 Rome, Italy}
\altaffiltext{3}{CEA Saclay, DSM/DAPNIA/Service d'Astrophysique (CNRS FRE
2591), F-91191, Gif sur Yvette, France}

\shorttitle{Magnetars and SFXTs}
\shortauthors{E. Bozzo, et al.}

\begin{abstract}
In this paper we survey the theory of wind accretion in high mass X-ray binaries 
hosting a magnetic neutron star and a supergiant companion.
We concentrate on the different types of interaction between the inflowing 
wind matter and the neutron star magnetosphere that are relevant when 
accretion of matter onto the neutron star surface is largely inhibited;  
these include the inhibition through the centrifugal and magnetic barriers.  
Expanding on earlier work, we calculate the expected luminosity for each regime and 
derive the conditions under which transition from one regime to another can take place. 
We show that very large luminosity swings ($\sim$10$^{4}$ or more
on time scales as short as hours) can result from transitions across different 
regimes. 
The activity displayed by supergiant fast X-ray transients, 
a recently discovered class of high mass X-ray binaries in our galaxy,
has often been interpreted in terms of direct accretion onto a neutron star
immersed in an extremely clumpy stellar wind. 
We show here that the transitions across the magnetic and/or centrifugal barriers 
can explain the variability properties of these sources as a results of 
relatively modest variations in the stellar wind velocity and/or density. 
According to this interpretation we expect that supergiant fast X-ray transients 
which display very large luminosity swings and host a slowly spinning neutron star 
are characterized by magnetar-like fields, irrespective of whether the magnetic 
or the centrifugal barrier applies. Supergiant fast X-ray transients might 
thus provide a new opportunity to detect and study magnetars in binary systems. 
\end{abstract}

\keywords{accretion, accretion disks --- stars: neutron --- supergiant --- X-rays: binaries --- X-rays: stars}

\section{Introduction}
\label{sec:intro}

High mass X-ray binaries (HMXBs) consist of a collapsed object,
usually a magnetic neutron star (NS), that accretes matter from an OB companion star. 
Mass transfer takes place because of the intense stellar wind from the OB 
star, part of which is captured by the collapsed object \citep[e.g.][]{verbunt}. 
Only in some short orbital period systems, the early type star, often a 
supergiant, fills its Roche lobe and leads to mass transfer through 
Roche lobe overflow \citep{tauris}. Persistent HMXBs accrete all the time and in most 
cases display X-ray luminosities in the 10$^{35}$-10$^{38}$~erg~s$^{-1}$ range. 
Many HMXBs are transient systems that remain at low 
X-ray luminosity levels (10$^{32}$-10$^{33}$~erg~s$^{-1}$) most of the time 
and undergo outbursts lasting from weeks to months. 
During these outbursts they display nearly identical properties 
to those of persistent HMXBs. 
Transient systems usually comprise a Be star donor and relatively 
long, moderately eccentric orbits, such that the star sits deep in its 
Roche lobe and stellar wind capture is the only mechanism through which 
mass transfer takes place. The occurrence of the outbursts is likely 
associated to variations in the stellar wind of the Be star, such 
as shell ejection episodes, or build up of matter around the 
resonant orbits in the slow equatorial wind component \citep{heuvel}.   
However, there are characteristics of the outbursts that are difficult to 
interpret if accretion onto the neutron star surface takes place 
unimpeded also in quiescence; these are (a) the large 
outburst to quiescence X-ray luminosity swing (factor of $\sim$10$^3$ or 
larger) and (b) the presence in
a given source of low-luminosity (Type I) outbursts recurring close to 
periastron and, at different times, of high-luminosity (Type II) outbursts 
that last for several orbital cycles and display little (if any) X-ray flux 
variations associated to the orbital phase. These characteristics of 
Be transients can be explained if the accretion rate (and thus X-ray 
luminosity) variations that are produced by the stellar wind alone, could 
be amplified by some ``gating'' mechanism. Since most Be star HMXB transients
contain relatively fast spinning X-ray pulsars, such mechanism has been 
identified with the centrifugal barrier that 
results from the rotation of the neutron star magnetosphere \citep{stella86}. 

About 10 transient systems have been recently 
discovered, which display sporadic outbursts lasting from 
minutes to hours (i.e. much shorter than Be star transients') 
and reach peak luminosities of $\sim$10$^{36}$-10$^{37}$~erg~s$^{-1}$.   
These systems spend long time intervals in quiescence, with X-ray 
luminosities down to $\sim$10$^{-5}$ times lower than those in outburst; 
in spite of their association with OB supergiant companions, 
their behaviour is thus at variance with other persistent 
and transient HMXBs. They define a new class of HMXBs, collectively 
termed supergiant fast X-ray transients, SFXTs.
An overview of the properties of SFXTs is given in
\S~\ref{sec:properties}.

If accretion onto the collapsed object of SFXTs takes place both 
in quiescence and outburst, then the corresponding X-ray luminosity 
swing, typically a factor of $\sim$10$^4$-10$^5$, would require  
wind inhomogeneities
with a very large density and/or velocity contrast   
\citep[according to the standard wind accretion, the mass capture 
rate onto the NS scales like $\dot{M}_{\rm w}$v$_{\rm w}^{-4}$, 
with $\dot{M}_{\rm w}$, the mass loss rate
and v$_{\rm w}^{-4}$, the wind velocity of the supergiant star,][]{davidson}. 
Several authors \citep{zand05,leyder,walter07} 
suggested the presence of dense clumps in the wind of the OB 
companions in order to attain the luminosity variations of SFXTs. 
While some observations provide evidence for a clumpy wind, the 
characteristics of such inhomogeneities are still poorly known. 
Numerical simulations suggest that clumps may originate from small 
scale perturbations in the radiation-driven wind \citep{dessart03,prinja}. 
Models involving accretion of clumps are still being 
actively pursued \citep{negueruela08,walter07}.   
The requirement on the density and/or velocity contrasts 
in the wind can be eased if there is a barrier
that remains closed during quiescence, halting
most of the accretion flow, and opens up 
in outbursts, leading to direct accretion 
(this is similar to the case of Be star transients). 
This paper develops and discusses gated accretion models for SFXTs.  

In \S~\ref{sec:model} after reviewing the 
theory of wind accretion in HMXBs, we describe the different regimes 
of a rotating magnetic neutron star immersed the stellar wind 
from its companion. 
In \S~\ref{sec:trans} we 
discuss the conditions under which transitions across
regimes can take place in response to variations in the wind parameters. 
As compared to previous works addressing gating mechanisms
in transient accreting magnetic neutron stars \citep[notably][]{stella86}, 
we present here a more comprehensive treatment of the different physical 
processes that have been discussed
in this context. 

In \S~\ref{sec:results}, we present an application  
to the different states of two 
SFXTs and discuss in turn the possibility that the 
outbursts are driven by a 
centrifugal or a magnetic barrier. 
The latter mechanism involves a magnetosphere 
extending beyond the accretion radius \citep{bondi},  
which prevents most of the inflowing matter from accreting; 
it is discussed here for the first time in relation to the activity 
of transient binary sources. 
Unless the stellar wind were extremely clumpy (as envisaged in
other SFXT models), the onset of a barrier inhibiting direct accretion
would be required to explain the activity of SFXTs. In this case  
we conclude that if SFXTs host slowly rotating neutron stars 
(spin periods of several hundreds to thousands seconds), then they 
must possess magnetar-like fields 
($\sim$10$^{14}$-10$^{15}$~G), independent of whether the centrifugal 
or magnetic barrier operates.  
We summarize our discussion and conclusions in \S~\ref{sec:discussion} 
and \S~\ref{sec:conclusions}. 

\begin{figure*}[t!]
\centering
\includegraphics[height=5.4 cm]{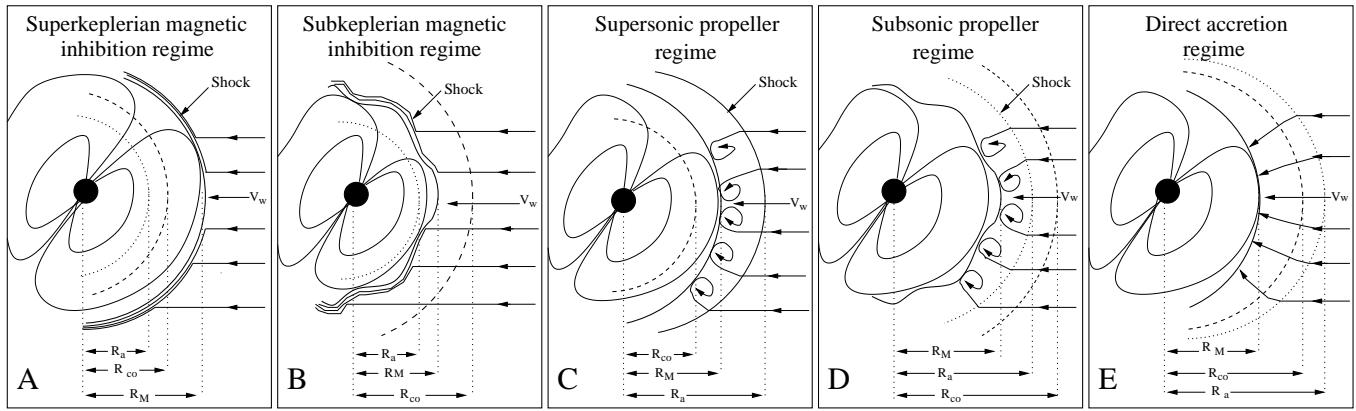}
\smallskip
\caption{Schematic view of a magnetized NS interacting with the 
inflowing matter from its supergiant companion. All the regimes of 
\S~\ref{sec:model} are shown, together with the relative 
position of the magnetospheric 
radius (solid line), the corotation radius (dashed line), and the accretion 
radius (dotted line). A wavy solid line is used when the magnetospheric 
boundary at R$_{\rm M}$ is Kelvin-Helmholtz unstable. In the supersonic and subsonic 
propeller regime convective motions at the base of the atmosphere are 
represented with small eddies.}
\label{fig:model} 
\end{figure*}

\section{The observed properties of SFXTs}
\label{sec:properties}

SFXTs are observed to exhibit sporadic outbursts, lasting from minutes to hours, 
with peak X-ray luminosities between $\sim$10$^{36}$ and 10$^{37}$~erg~s$^{-1}$ 
\citep[see e.g.,][]{gon04,sidoli05,grebenev05,Lutovinov05,sguera05,masetti06,
sguera06,gotz07,sguera07}. 
No firm orbital period measurement  
has been obtained yet\footnote{Only \objectname{IGR J11215-5952} and \objectname{AX J1749.1-2733}   
showed recurrent flaring activity, with periodicity of $\sim$165~d and $\sim$185~d, respectively. 
These are interpreted as outbursts from two systems with unusually long orbital 
periods \citep[$\gtrsim$100~d,][]{grebenev05,sidoli07,zurita07}. 
Thus \citet{walter07} excluded these sources from their SFXT list.}. 
A recent list of confirmed ($\sim$5) and candidate ($\sim$6)  
SFXTs is given by \citet{walter07}. 

Between outbursts, SFXTs remain in quiescence with 
luminosities in the range $\sim$10$^{31}$-10$^{33}$~erg~s$^{-1}$  
\citep{gon04,zand05,smith06,kennea06}. 
In some cases, very high peak-to-quiescence X-ray luminosity swings 
(factor of $\sim$10$^{4}$-10$^{5}$) were seen    
on timescales comparable to the outburst duration. 
Some SFXTs showed also flare-like
activity at intermediate luminosity levels \citep[e.g.,][]{gon04}.  
In the case of \objectname{IGR J17544-2619}, two states of intermediate luminosity 
were observed: one before the onset of the outburst and the other immediately after,  
with X-ray luminosities $\sim$3 and $\sim$1 decades below 
the value reached at the peak of the outburst, respectively \citep{zand05}. 

The sporadic character of SFXT outbursts, as observed with \int,\ suggested that the 
duty cycle of these sources (the fraction of time spent in a high luminosity state) 
is small \citep[$\sim$0.02-0.002,][]{walter07}. 
However, recent observations carried out with the very sensitive X-ray telescopes on board 
\xmm\ and \chan\ revealed that some SFXTs display flares around a luminosity of 
$\gtrsim$10$^{34}$-10$^{35}$~erg~s$^{-1}$ (i.e. well below the \int\ 
limiting sensitivity) for a large fraction of the time
\citep{gon04,zand05,tomsick06}.  
Therefore, the indication is that the active phase of SFXT sources (as opposed to true
quiescence) lasts longer than previously thought, and the duty cycles are of order 
$\sim$0.1 or higher.

Optical identifications of SFXTs show that these sources are
associated to OB supergiant companion stars \citep[see e.g.,][and 
reference therein]{walter07}. The SFXT 
OB companions have typically mass of   
M$_{*}$$\sim$30M$_{\rm \odot}$, luminosity of 
log (L$_{*}$/L$_{\odot}$)$\sim$5-6, mass loss 
rate of $\dot{M}_{\rm w}$=10$^{-7}$-10$^{-5}$~M$_{\odot}$~yr$^{-1}$,  
and wind velocity of 1000-2000~km~s$^{-1}$. 
Note, that isolated OB stars with log(L$_{*}$/L$_{\odot}$)$\sim$5-6 
are persistent soft X-ray sources with luminosity 
around $\sim$10$^{32}$~erg~s$^{-1}$ \citep{cassinelli81,berghofer97}.  

It is widely believed that SFXTs contain sporadically accreting 
neutron stars. Only little is known about their spin period.  
A coherent periodicity was detected at 4.7~s and  228~s 
in \objectname{AX J1841.0-0536} and \objectname{IGR J16465-4507},  
respectively \citep{bamba01,Lutovinov05}. However, \igrjj\  
showed only a factor of $\sim$100 luminosity swing 
between quiescence and outburst, so it is unclear whether the source 
should be considered a transient  
\citep[in fact it is classified as an ``intermediate system'',]
[]{walter07}. The nature of the companion star in \axj\ is 
still debated \citep{halpern04,nespoli07}. 
Therefore, these two sources might not belong 
to the SFXT class. 
On the contrary, in the prototypical SFXTs 
\objectname{XTE J1739-302} \citep{sguera06} and \objectname{IGR J16479-4514} \citep{walter06} 
some evidence has been reported for periodicities in the 
$\sim$1000-2000~s range. We assume in the following that SFXTs host a  
rotating magnetic neutron star.

\section{Stellar wind accretion}
\label{sec:model} 

We investigate here the conditions under which a magnetized 
neutron star can accrete matter from the 
wind of a massive companion. 
In the theory of wind accretion in HMXBs, the following radii are defined
\citep[see e.g.,][]{illarionov75, stella86}: 
\begin{itemize}

\item The accretion radius, R$_{\rm a}$ is the distance at which the 
inflowing matter is gravitationally focused toward the NS \citep{bondi}. 
It is usually expressed as   
\begin{equation}
R_{\rm a}=2GM_{\rm NS}/v_{\rm w}^2= 3.7\times10^{10} v_{8}^{-2} ~{\rm cm},
 \label{eq:ra}
\end{equation}
where v$_{8}$ is the wind velocity in units of 1000~km~s$^{-1}$ and we 
assumed that the orbital velocity of the star 
is negligible \citep{fkr}. Throughout the paper we fix the NS radius and 
mass at R$_{\rm NS}$=10$^6$~cm and M$_{\rm NS}$=1.4~M$_\odot$, respectively. 
The fraction $\dot{M}_{\rm capt}$/$\dot{M}_{\rm w}$ of the stellar wind mass loss rate 
($\dot{M}_{\rm w}$) captured by the NS depends on R$_{\rm a}$ through 
\citep{fkr} 
\begin{equation}
\dot{M}_{\rm capt}/\dot{M}_{\rm w}\simeq R_{\rm a}^2/(4 a^2)=2\times10^{-5} v_{8}^{-4} 
a_{\rm 10d}^{-2}. 
\label{eq:dotmcapt}
\end{equation}
Here a=4.2$\times$10$^{12}$a$_{\rm 10d}$~cm is the orbital separation, 
a$_{\rm 10d}$=P$_{\rm 10d}^{2/3}$M$_{30}^{1/3}$, P$_{\rm 10d}$ is the binary orbital 
period in units of 10 days, and M$_{30}$ is the total 
mass in units of 30~M$_{\odot}$ (we assumed circular orbits). 
 
\item The magnetospheric radius, R$_{\rm M}$, at which the pressure of the NS magnetic 
field ($\mu^2$/(8$\pi$$R_{\rm NS}^6$), with $\mu$ the NS magnetic moment) 
balances the ram pressure of the inflowing matter 
($\rho_{\rm w}$$v_{w}^{2}$). 
In the case in which R$_{\rm M}$$>$R$_{\rm a}$, 
the magnetospheric radius is given by \citep{pringle}  
\begin{equation}
R_{\rm M}=3.3\times10^{10} \dot{M}_{-6}^{-1/6} v_{8}^{-1/6} a_{\rm 10d}^{1/3} 
\mu_{33}^{1/3} ~ {\rm cm}.   
\label{eq:rm}
\end{equation} 
Here we assumed a non magnetized spherically symmetric wind \citep{elsner1977},  
with density\footnote{We approximated a-R$_{\rm M}$$\simeq$a ,  
which is satisfied for a very wide range of parameters.} 
$\rho_{\rm w}$(R$_{\rm M}$)$\sim$$\dot{M}_{\rm w}$/(4$\pi$a$^2$v$_{\rm w}$), 
a dipolar NS magnetic field with $\mu_{33}$=$\mu$/10$^{33}$~G~cm$^{3}$, and 
$\dot{M}_{-6}$=$\dot{M}_{\rm w}$/10$^{-6}$ M$_{\odot}$~yr$^{-1}$. 
In the following sections we discuss the range of applicability of Eq.~\ref{eq:rm}, 
and the regimes in which a different prescription for R$_{\rm M}$ should be 
used.  

\item The corotation radius, R$_{\rm co}$, at which the NS angular velocity 
equals the Keplerian angular velocity, i.e. 
\begin{equation}
R_{\rm co}=1.7\times10^{10} P_{\rm s3}^{2/3} ~{\rm cm}.    
\label{eq:rco}
\end{equation}
Here $P_{\rm s3}$ is the NS spin period in units of 10$^3$~s. 
\end{itemize}

Changes in the relative position of these radii
result into transitions across different regimes for the NS 
\citep{illarionov75, stella86}. Below we discuss these regimes 
singularly, and provide a schematic representation 
of each regime in Fig.~\ref{fig:model}. 
Being determined primarily by the spin 
of the neutron star, the corotation radius can change only over 
evolutionary timescales. \citet{illarionov75} summarises 
the different regimes experienced by a spinning down NS since
its birth, from the initial radio pulsar (i.e. rotation powered) stage, 
to the regime in which mass accretion onto the neutron star surface can 
take place. On the other hand the accretion radius and magnetospheric 
radius depend on the wind parameters (see Eqs.~\ref{eq:ra} and \ref{eq:rm}),
which can vary on a wide range of timescales (from hours to months).  
Therefore, variations in the wind parameters can cause the neutron 
star to undergo transitions across different regimes on comparably 
short timescales, thus opening the possibility to explain the 
properties of some classes of highly variable X-ray sources through them. 
In particular, the transition from 
the accretion regime to the propeller regime (and vice versa), across the 
so-called ``centrifugal barrier'', was identified as a likely 
mechanism responsible for the pronounced activity of Be X-ray pulsar 
transient systems \citep{stella86}.  
Below we summarise the different regimes of a 
magnetic rotating neutron star, subject to a varying 
stellar wind, with special attention to the condition under 
which a ``magnetic'' (as opposed to ``centrifugal'')
barrier inhibits accretion onto the neutron star
\citep{hard, ruth, mori, toropina, toropina2}. 
As it will be clear in the following, new motivation for 
investigating magnetic inhibition of accretion comes from 
the discovery of magnetars, neutrons stars with 
extremely high magnetic fields \citep[$\sim$10$^{14}$-10$^{15}$~G,][]{dun}.

\subsection{Outside the accretion radius: the magnetic inhibition of accretion: 
R$_{\rm M}$$>$R$_{\rm a}$}

We consider here the case in which the magnetospheric radius is larger than 
the accretion radius\footnote{A similar case was considered  
also by \citet{lipunov02}, ``the georotator regime'', and by \citet{toropina2},  
``the magnetic plow regime''.}.  
In systems with R$_{\rm M}$$>$R$_{\rm a}$ the mass flow from the companion star interacts 
directly with the NS magnetosphere without significant gravitational focusing, forming  
a bow shock at R$_{\rm M}$ \citep{hard, toropina2}. 
A region of shocked gas surrounds the NS magnetosphere with density 
$\rho_{\rm ps}$$\simeq$4$\rho_{\rm w}$ and velocity 
v$_{\rm ps}$$\simeq$v$_{\rm w}$/4 (the subscript ``ps'' stands for 
post-shock). These are only rough estimates because 
the shock is very close to the magnetopause and it does not satisfy the 
standard Rankine-Hugoniot conditions \citep{toropina2}. At least 
in the front part of the shock, i.e. in the region around the  
stagnation point, the whole kinetic energy of the inflowing matter is    
converted into thermal energy, and the expected temperature of the heated gas is  
T$\simeq$m$_{\rm p}$v$_{\rm w}^2$/(3k)$\simeq$4$\times$10$^7$v$_{8}^2$~K. 
Thus, the power released in this region is of order  
\begin{equation}
L_{\rm shock}\simeq\frac{\pi}{2} R_{\rm M}^2 \rho_{\rm w} v_{\rm w}^3 = 
4.7\times10^{29} R_{\rm M10}^2 
v_8^2 a_{\rm 10d}^{-2} \dot{M}_{-6}~{\rm erg ~s^{-1}}  
\label{eq:lshock}
\end{equation}
(R$_{\rm M10}$ is the magnetospheric radius in units of 10$^{10}$~cm), 
and is mainly radiated in the X-ray band \citep{toropina}. 
Below we distinguish two different regimes of magnetic inhibition of accretion.  

\subsubsection{The superKeplerian magnetic inhibition regime: R$_{\rm M}$$>$R$_{\rm a}$, 
R$_{\rm co}$}
\label{sec:mprop} 
  
In this ``superKeplerian'' magnetic inhibition regime the magnetospheric radius is larger than 
both the accretion and corotation radii (R$_{\rm M}$$>$R$_{\rm a}$, R$_{\rm co}$).
Matter that is shocked and halted close to R$_{\rm M}$ cannot 
proceed further inward, due to the rotational drag of the NS 
magnetosphere which is locally superKeplerian. 
Since magnetospheric rotation is also supersonic\footnote{This can be easily 
seen by comparing v$_{\rm w}$ with $\Omega$R$_{\rm M}$, for the 
value of the parameters used in this section.}, the interaction 
between the NS magnetic field and matter at R$_{\rm M}$ results in 
rotational energy dissipation and thus, NS spin down. 
In order to derive an upper limit on the contribution of this dissipation to the overall 
luminosity, we assume that the above interaction is anelastic \citep[e.g.,][]{perna06}, 
i.e. that matter at R$_{\rm M}$ is forced to corotate. This process releases energy at 
a rate 
\begin{equation}
L_{\rm sd} \simeq \pi R_{\rm M}^2 \rho_{\rm w} v_{\rm w} (R_{\rm M}\Omega)^2\simeq  
3.7\times10^{29} R_{\rm M10}^4 \dot{M}_{-6} a_{\rm 10d}^{-2} P_{\rm s3}^{-2}~{\rm erg ~s^{-1}}, 
\label{eq:lmagnsd}
\end{equation}
which adds to the shock luminosity (Eq.~\ref{eq:lshock}).

\subsubsection{The subKeplerian magnetic inhibition regime: R$_{\rm a}$$<$R$_{\rm M}$$<$R$_{\rm co}$}
\label{sec:intermediate}

If R$_{\rm a}$$<$R$_{\rm M}$$<$R$_{\rm co}$ the magnetospheric drag is subKeplerian
and matter can penetrate the NS magnetosphere. 
In this ``subKeplerian'' magnetic inhibition regime, the boundary between  
the inflowing matter and the magnetosphere is subject to the Kelvin-Helmholtz 
instability \citep[KHI,][]{hard}. The mass inflow rate across R$_{\rm M}$ resulting
from the KHI is approximately \citep{bur} 
\begin{eqnarray}
\dot{M}_{\rm KH}\simeq2\pi R_{\rm M}^2 \rho_{\rm ps} v_{\rm conv} = 
2 \pi R_{\rm M}^2 \rho_{\rm ps} v_{\rm sh} 
 \eta_{\rm KH} (\rho_{\rm i}/\rho_{\rm e})^{1/2}  
(1+\rho_{\rm i}/\rho_{\rm e})^{-1},  
\label{eq:dotmkh}
\end{eqnarray} 
where $\eta_{\rm KH}$$\sim$0.1 is an efficiency factor,  
v$_{\rm sh}$ is the shear velocity, $\rho_{\rm i}$ and $\rho_{\rm e}$ the density inside 
and outside the magnetospheric boundary at R$_{\rm M}$, respectively.  
Close to the stagnation point, virtually all the kinetic energy 
of the wind matter is converted into thermal energy (see also \S~\ref{sec:mprop}), and 
the shear velocity is thus dominated by the magnetosphere's rotation, 
such that v$_{\rm sh}$=v$_{\rm rot}$=2$\pi$P$_{\rm s}^{-1}$R$_{\rm M}$. 
Away from this region, the tangential component of the wind velocity with respect 
to the NS magnetic field lines increases up to v$_{\rm ps}$ and, in general, 
the shear velocity and rate at which plasma enters the magnetosphere 
due to the KHI, depends on both v$_{\rm ps}$ and v$_{\rm rot}$. 
For the aims of this paper, we adopt v$_{\rm sh}$=max(v$_{\rm ps}$, v$_{\rm rot}$), 
and use mass conservation across the KHI unstable layer to estimate the 
density ratio at R$_{\rm M}$ \citep{bur}. If matter crossing the unstable layer of height h$_{\rm t}$ 
is rapidly brought into corotation with the NS magnetosphere and free-falls onto 
the NS, mass conservation implies 
\begin{equation}
R_{\rm M}^2 \rho_{\rm e} v_{\rm conv}\simeq R_{\rm M} h_{\rm t} \rho_{\rm i} v_{\rm ff}(R_{\rm M}). 
\label{eq:masscons}
\end{equation} 
The height h$_{\rm t}$ of the unstable layer, where matter and magnetic field coexist, is mostly 
determined by the largest wavelength of the KHI unstable mode \citep{bur}. 
A detailed analysis of this instability is beyond the scope of this paper. 
In Eq.~\ref{eq:masscons} we use conservatively h$_{\rm t}$$\simeq$R$_{\rm M}$, 
and discuss in Appendix~\ref{app:ht} the effect of smaller values of h$_{\rm t}$.   
Therefore accretion of matter at a rate $\dot{M}_{\rm KH}$ onto the NS is  
expected to release a luminosity of  
\begin{eqnarray}
L_{\rm KH} = 3.5\times10^{34} \eta_{\rm KH} R_{\rm M10}^2 a_{\rm 10d}^{-2} \dot{M}_{-6} 
(\rho_{\rm i}/\rho_{\rm e})^{1/2} (1+\rho_{\rm i}/\rho_{\rm e})^{-1}~{\rm erg~s^{-1}}, 
\label{eq:lx_kh}
\end{eqnarray}
if v$_{\rm sh}$=v$_{\rm ps}$, or 
\begin{eqnarray}
L_{\rm KH} = 8.8\times10^{34} \eta_{\rm KH} P_{\rm s3}^{-1} R_{\rm M10}^3 
a_{\rm 10d}^{-2} v_8^{-1} \dot{M}_{-6} 
(\rho_{\rm i}/\rho_{\rm e})^{1/2} (1+\rho_{\rm i}/\rho_{\rm e})^{-1} ~{\rm erg~s^{-1}}, 
\label{eq:lx_kh2}
\end{eqnarray}
if v$_{\rm sh}$=v$_{\rm rot}$. 
The values of $\rho_{\rm i}$/$\rho_{\rm e}$ that we use in Eqs.~\ref{eq:lx_kh} and \ref{eq:lx_kh2} 
are derived numerically from Eq.~\ref{eq:masscons}. 

In the subKeplerian magnetic inhibition regime, plasma penetration inside the 
NS magnetosphere is sustained also by Bohm diffusion \citep{ikhsanov}. 
This diffusion, being dependent on the temperature of the plasma (rather than velocity), 
is highest in the region close to the stagnation point, where the shock slows down 
the inflowing plasma most efficiently (see also \S~\ref{sec:mprop}). 
In accordance with \citet{ikhsanov2001}, the maximum inflow rate  
allowed by Bohm diffusion is 
\begin{eqnarray}
\dot{M}_{\rm diff} \simeq 2\pi R_{\rm M}^2 \rho_{\rm ps} V_{\rm m} = 
4.5\times10^{9} \zeta^{1/2} \dot{M}_{-6} \mu_{33}^{-1/2} R_{\rm M10}^{11/4}  
a_{\rm 10d}^{-2} ~{\rm g ~s^{-1}}, 
\label{eq:dotmdiff} 
\end{eqnarray}
where V$_{\rm m}$=$\sqrt{D_{\rm eff}/t_{\rm ff}}$ is the diffusion velocity, 
D$_{\rm eff}$=($\zeta$ckT$_{\rm i}(R_{\rm M})$)/(16eB(R$_{\rm M}$)) the diffusion coefficient, 
T$_{\rm i}$(R$_{\rm M}$)$\simeq$m$_{\rm p}$v$_{\rm w}^2$/(3k) the post-shock ion temperature, 
t$_{\rm ff}$=$\sqrt{R_{\rm M}^3/(2GM)}$ the free-fall time, $\zeta$$\simeq$0.1  
an efficiency factor and m$_{\rm p}$ the proton mass. In the above equation we also 
approximated the density outside R$_{\rm M}$, around the stagnation point, with 
$\rho_{\rm ps}$ \citep[though this might be underestimate by a factor of a few, see][]{toropina2}.  
Over the whole range of parameters relevant to this work, 
the diffusion-induced mass accretion rate
is orders of magnitude smaller than that due to the KHI. 
Similarly, the contribution to the total luminosity resulting from
the shock and anelastic drag at the magnetospheric boundary can be 
neglected in this regime.

\subsection{Inside the accretion radius: R$_{\rm M}$$<$R$_{\rm a}$}

\subsubsection{The supersonic propeller regime: R$_{\rm co}$$<$R$_{\rm M}$$<$R$_{\rm a}$}
\label{sec:supersonic}

Once R$_{\rm M}$ is inside the accretion radius, matter flowing from the 
companion star is shocked adiabatically at R$_{\rm a}$ and halted at the NS 
magnetosphere. In the region between R$_{\rm a}$ and R$_{\rm M}$  
this matter redistributes itself into an approximately spherical
configuration  
(resembling an ``atmosphere''), whose shape and properties 
are determined by the interaction between matter and 
NS magnetic field at R$_{\rm M}$. 
This scenario was considered previously by \citet{davies} 
and \citet{pringle}, and we follow here their treatment. 
These authors demonstrated that hydrostatic equilibrium ensues 
when radiative losses inside 
R$_{\rm a}$ are negligible (we discuss this approximation 
in Appendix~\ref{app:supersonic} and \ref{app:subsonic}) and the atmosphere is 
stationary on dynamical time-scales.  
Assuming a polytropic law of the form p$\propto$$\rho^{1+1/n}$, 
the pressure and density of this atmosphere are:
\begin{eqnarray}
p(R)=\rho_{\rm ps} v_{\rm ps}^2 \left[1+(1/(1+n))8R_{\rm a}/R\right]^{n+1} \\
\rho(R)=\rho_{\rm ps}\left[1+(1/(1+n))8R_{\rm a}/R\right]^{n}. ~~~~~~ 
\label{eq:hydro} 
\end{eqnarray} 
The value of the polytropic index $n$ depends on the conditions at 
the inner boundary of the atmosphere, and in particular on the rate 
at which energy is deposited there.  

When the rotational velocity of the NS magnetosphere at R$_{\rm M}$ 
is supersonic (see also \S~\ref{sec:mprop}), 
the interaction with matter in the atmosphere leads  
to dissipation of some of the star's rotational energy and thus spin-down. 
In the supersonic propeller regime, \citet{pringle} showed that 
turbulent motions are generated at R$_{\rm M}$ which convect this 
energy up through the atmosphere, until it is lost 
at its outer boundary. In this case $n$=1/2. 
Accordingly, taking into account the structure of the surrounding atmosphere,
the magnetospheric radius is given by  
\begin{equation}
R_{\rm M}^{-6}(1+16 R_{\rm a}/ (3 R_{\rm M}))^{-3/2}=\frac{1}{2}\dot{M}_{\rm w} a^{-2} 
\mu^{-2} v_{\rm w}. 
\label{eq:rmsuper} 
\end{equation} 
This can be approximated by
\begin{equation}
R_{\rm M}\simeq2.3\times10^{10} a_{\rm 10d}^{4/9} \dot{M}_{-6}^{-2/9} 
v_8^{4/9} \mu_{33}^{4/9} ~{\rm cm} .
\label{eq:rmsuperapprox} 
\end{equation}
Matter that is shocked at $\sim$R$_{\rm a}$, reaches the magnetospheric boundary at 
R$_{\rm M}$ where the interaction with the 
NS magnetic field draws energy from NS rotation 
(see also \S~\ref{sec:mprop}). According to \citet{pringle}, this 
contributes     
\begin{equation}
L_{\rm sd} = 2\pi R_{\rm M}^2 \rho(R_{\rm M}) c_{\rm s}^3
(R_{\rm M})\simeq5.4\times10^{31} \dot{M}_{-6}
a_{\rm 10d}^{-2} v_8^{-1} R_{\rm M10}^{1/2} 
(1+16 R_{\rm a10}/(3 R_{\rm M10}))^{1/2} ~{\rm erg ~s}^{-1} 
\label{eq:lspsuper}
\end{equation}
to the total luminosity.
In the above equation R$_{\rm a10}$=10$^{-10}$R$_{\rm a}$
and c$_{\rm s}$(R$_{\rm M}$)=v$_{\rm ff}$ 
(R$_{\rm M}$)=(2GM$_{\rm NS}$/R$_{\rm M}$)$^{1/2}$ \citep{pringle}. 
In the supersonic propeller regime the energy released through 
the shock at R$_{\rm a}$, 
\begin{equation}
L_{\rm shock} =\frac{9}{32}\pi R_{\rm a}^2 \rho_{\rm w} v_{\rm w}^3  
\simeq 2.6\times10^{29} R_{\rm a10}^2 
v_8^2 a_{\rm 10d}^{-2} \dot{M}_{-6} ~{\rm erg ~s^{-1}},   
\label{eq:lshock2}
\end{equation}  
is negligible.

\subsubsection{The subsonic propeller regime: R$_{\rm M}$$<$R$_{\rm a}$, R$_{\rm co}$, 
              $\dot{M}_{\rm w}$$<$$\dot{M}_{\rm lim}$}
\label{sec:subsonic}

The break down of the supersonic propeller regime occurs when 
R$_{\rm M}$$<$R$_{\rm co}$, i.e., when the magnetosphere rotation 
is no longer supersonic with respect to the 
surrounding material. The structure of the atmosphere changes and   
the transition to the subsonic propeller regime takes place. 
Since the rotation of the magnetosphere is subsonic, the atmosphere is roughly 
adiabatic ($n$=3/2), and the magnetospheric radius is given by 
\citep{pringle}:
\begin{equation}
R_{\rm M}^{-6}(1+16 R_{\rm a}/(5 R_{\rm M}))^{-5/2}=
\frac{1}{2}\dot{M}_{\rm w} a^{-2} \mu^{-2} v_{\rm w}. 
\label{eq:rmsub}
\end{equation} 
This can be approximated by
\begin{equation}
R_{\rm M}\simeq2\times10^{10} a_{\rm 10d}^{4/7} 
\dot{M}_{-6}^{-2/7} v_8^{8/7} \mu_{33}^{4/7} ~{\rm cm}. 
\label{eq:rmsubapprox} 
\end{equation}
In the subsonic propeller regime, the centrifugal barrier 
does not operate because R$_{\rm M}$$<$R$_{\rm co}$,  
but the energy input at the base of the 
atmosphere (due to NS rotational 
energy dissipation) is still too high for matter 
to penetrate the magnetosphere at a rate $\dot{M}_{\rm capt}$ 
\citep{pringle}. 
Nevertheless a fraction of the 
matter inflow at R$_{\rm a}$  is expected to accrete onto the 
NS, due to the KHI and Bohm diffusion\footnote{To our knowledge this is the first 
application of the KHI to the subsonic propeller regime.}.  

Based on the discussion 
in \S~\ref{sec:intermediate}, we estimate the accretion luminosity 
of this matter by using Eqs.~\ref{eq:dotmkh} and 
\ref{eq:dotmdiff} (we approximate here the surface of interaction  
between matter and magnetic field with 4$\pi$R$_{\rm M}^2$). This gives   
\begin{equation}
L_{\rm diff} \simeq G M_{\rm NS} \dot{M}_{\rm diff}/R_{\rm NS} 
=4.5\times10^{30} \dot{M}_{-6} a_{\rm 10d}^{-2}  
R_{\rm M10}^{9/4} \mu_{33}^{-1/2} \zeta^{1/2} v_8^{-1} 
(1+16 R_{\rm a10}/(5 R_{\rm M10}))^{3/2} ~{\rm erg ~s}^{-1},  
\label{eq:ldiffsub}
\end{equation}
and   
\begin{eqnarray}
& & L_{\rm KH} \simeq G M_{\rm NS} \dot{M}_{\rm KH}/R_{\rm NS}=\nonumber \\ 
& & 1.8\times10^{35} \eta_{\rm KH} P_{\rm s3}^{-1} R_{\rm M10}^3
\dot{M}_{-6} a_{\rm 10d}^{-2} v_8^{-1} 
(1+16 R_{\rm a10}/(5 R_{\rm M10}))^{3/2} 
(\rho_{\rm i}/\rho_{\rm e})^{1/2}
(1+\rho_{\rm i}/\rho_{\rm e})^{-1} ~{\rm erg ~s}^{-1} ,\nonumber \\
\label{eq:lkhsub} 
\end{eqnarray}
for the accretion luminosity arising from matter entering the  
magnetosphere through Bohm diffusion and KHI, respectively. 
For the range of parameters of interest here, 
Eqs.~\ref{eq:lshock}, \ref{eq:ldiffsub}, and \ref{eq:lkhsub} show that 
$L_{\rm KH}$ dominates. 
The rotational energy dissipation at R$_{\rm M}$ 
(see \S~\ref{sec:supersonic}) gives a small contribution with respect 
to Eq.~\ref{eq:lkhsub} \citep{pringle}:    
\begin{equation}
L_{\rm sd} =  2\pi R_{\rm M}^5 \rho(R_{\rm M}) \Omega^3 
= 2.2\times10^{30}  P_{\rm s3}^{-3} 
R_{\rm M10}^5 \dot{M}_{-6} v_8^{-1} a_{\rm 10d}^{-2} 
(1+16 R_{\rm a10}/(5 R_{\rm M10}))^{3/2} ~{\rm erg ~s}^{-1}.  
\label{eq:lspsub} 
\end{equation}

The subsonic propeller regime applies until the critical accretion rate 
\begin{equation}
\dot{M}_{\rm lim_{-6}} = 2.8\times10^2 P_{\rm s3}^{-3} 
a_{\rm 10d}^{2} v_8 R_{\rm M10}^{5/2} (1+16 R_{\rm a10}/(5 R_{\rm M10}))^{-3/2}
\label{eq:dotmlimsub} 
\end{equation}
is reached, at which the gas radiative cooling (bremsstralhung) completely damps  
convective motions inside the atmosphere (see Appendix~\ref{app:subsonic}). 
If this cooling takes place, direct accretion at a rate $\dot{M}_{\rm capt}$ 
onto the NS surface is permitted.

\subsubsection{The direct accretion regime: R$_{\rm M}$$<$R$_{\rm a}$, R$_{\rm co}$, 
               $\dot{M}_{\rm w}$$>$$\dot{M}_{\rm lim}$}
\label{sec:accretor}
If R$_{\rm M}$$<$R$_{\rm co}$ and matter outside the 
magnetosphere cools efficiently, 
accretion onto the NS takes place at the same rate  
$\dot{M}_{\rm capt}$ (see Eq.~\ref{eq:dotmcapt}) at which it 
flows towards the magnetosphere. 
The corresponding luminosity is 
\begin{equation}
L_{\rm acc}= G M_{\rm NS} \dot{M}_{\rm capt} / R_{\rm NS} = 
2\times10^{35} \dot{M}_{-6} a_{\rm 10d}^{-2} 
v_8^{-4} ~{\rm erg ~s}^{-1}\simeq2\times10^{35}\dot{M}_{15}~{\rm erg ~s}^{-1}, 
\label{eq:lacc} 
\end{equation}
where $\dot{M}_{15}$=$\dot{M}_{\rm capt}$/10$^{15}$ g s$^{-1}$. 
This is the standard accretion regime; the system achieves the highest mass to 
luminosity conversion efficiency.  

\begin{figure}
\centering
\includegraphics[height=22.0cm]{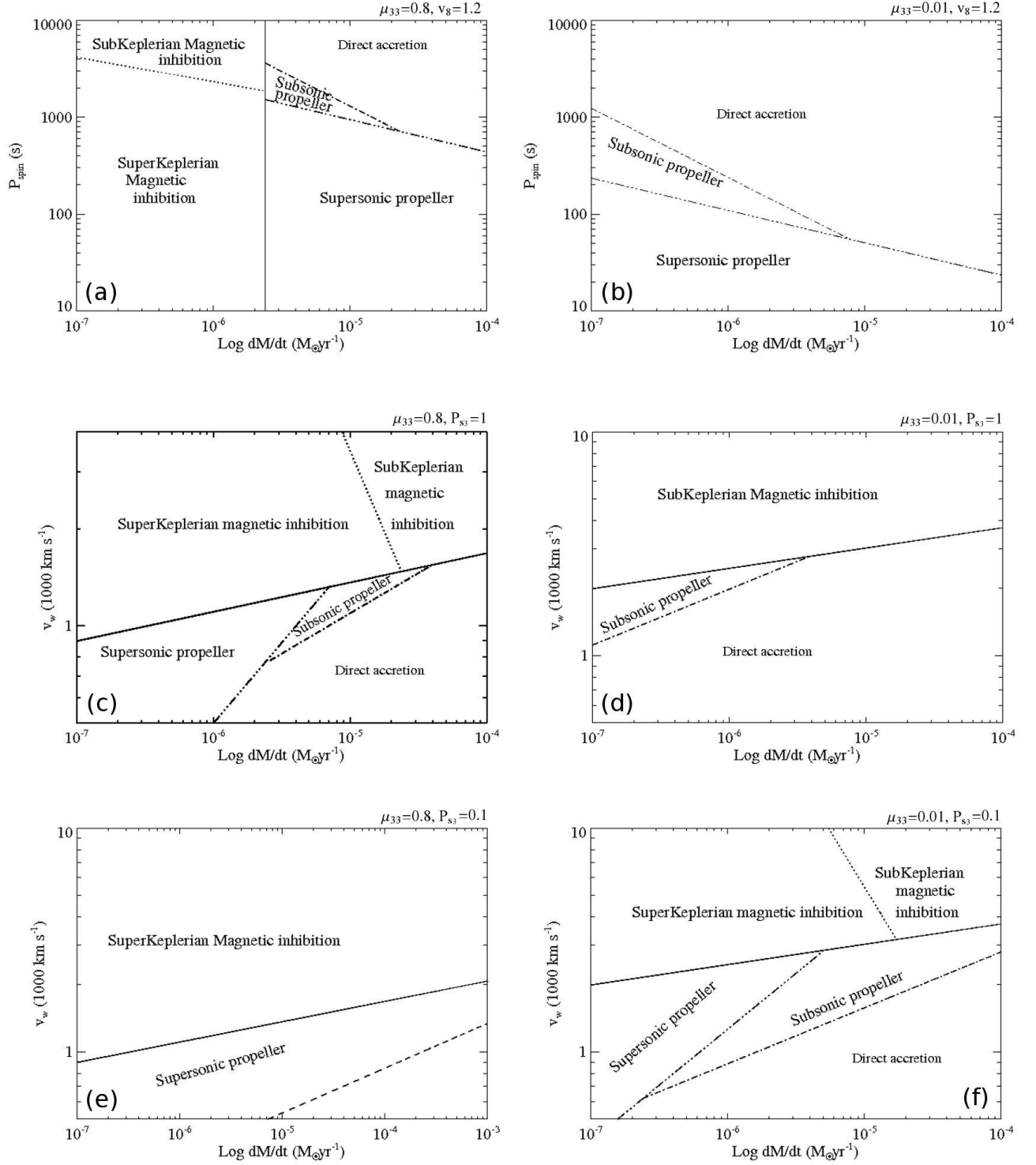}
\caption{Transitions between regimes described in \S~\ref{sec:model} 
for selected values of the parameters $\mu_{33}$, P$_{\rm s3}$, and v$_{8}$  
(in each panel these parameters are indicated in the top right corner). 
In all cases we fixed a$_{\rm 10d}$=1 (see \S~\ref{sec:trans}). 
In panels (a) and (b) we fixed $\mu_{33}$ and v$_{8}$ and investigated 
the different regimes in the  P$_{\rm s3}$-$\dot{M}_{-6}$ plane. 
In panels (c), (d), (e), and (f), instead, fixed parameters are P$_{\rm s3}$ 
and $\mu_{33}$, and the relevant regimes are shown in the 
v$_{8}$-$\dot{M}_{-6}$ plane. 
In all cases Eqs.~\ref{eq:dotmrarm}, \ref{eq:pspinlim}, \ref{eq:pspinlimsuper}, 
\ref{eq:pspinlimsub}, and \ref{eq:consistencysuper} (see Appendix~\ref{app:supersonic})   
are represented with a solid line, a dotted line, a dot-dot-dot dashed line, 
a dot-dashed line, and a dashed line, respectively. 
Note that the line from Eq.~\ref{eq:consistencysuper} is present only in panel (e), i.e  
our treatment of the the supersonic propeller is self-consistent everywhere 
except for a small region away from the range of interest for 
SFXT sources (see also \S~\ref{sec:results}).}   
\label{fig:trans}
\end{figure}

\section{Transitions and paths across different regimes}
\label{sec:trans}

We explore here the conditions under which 
transitions across different regimes take place. 
As emphasised in \S~\ref{sec:model}, these transitions occur 
when the relative positions of R$_{\rm M}$, R$_{\rm a}$, and R$_{\rm co}$
change; we concentrate here on transitions that occur in response to 
variations in the stellar wind parameters. In the following, 
since R$_{\rm M}$ depends only weakly on the orbital period and the 
total mass of the system, we fix a$_{\rm 10d}$=1 
(we explain this choice in \S~\ref{sec:results}), and 
investigate variations in the other four parameters: 
$\mu_{33}$, P$_{\rm s3}$, v$_{8}$, and $\dot{M}_{-6}$. 

The equations that define the conditions for transitions between different 
regimes are\footnote{Here we used 
Eqs.~\ref{eq:rmsuperapprox} and \ref{eq:rmsubapprox} for the magnetospheric 
radius in the supersonic and subsonic propeller regime, respectively.} 
\begin{equation}
R_{\rm M}>R_{\rm a} \Rightarrow \dot{M}_{-6}\lesssim0.45 \mu_{33}^2 v_{8}^{11} a_{\rm 10d}^2  
\label{eq:dotmrarm}
\end{equation} 
(or equivalently $\dot{M}_{15}$$\lesssim$0.6$\mu_{33}^2$v$_{8}^{7}$, see Eqs.~\ref{eq:ra}, 
\ref{eq:rm} and \ref{eq:dotmcapt}), for the magnetic barrier; 
\begin{equation} 
R_{\rm M}>R_{\rm co} \Rightarrow P_{\rm s3}\lesssim2.6\dot{M}_{-6}^{-1/4} 
v_{8}^{-1/4} a_{\rm 10d}^{1/2} \mu_{33}^{1/2}\simeq2.8\dot{M}_{15}^{-1/4}v_{8}^{-5/4}\mu_{33}^{1/2}  
\label{eq:pspinlim}
\end{equation}
(see Eq.~\ref{eq:rm}) if $R_{\rm M}>R_{\rm a}$, or 
\begin{equation}
R_{\rm M}>R_{\rm co} \Rightarrow P_{\rm s3}\lesssim1.8 a_{\rm 10d}^{2/3} 
\dot{M}_{-6}^{-1/3} v_{8}^{2/3} \mu_{33}^{2/3}\simeq2\dot{M}_{15}^{-1/3}v_{8}^{-2/3}\mu_{33}^{2/3} 
\label{eq:pspinlimsuper}
\end{equation}
(see Eq.~\ref{eq:rmsuperapprox}) if R$_{\rm M}$$<$R$_{\rm a}$, for the centrifugal barrier. 
 
The equation that defines the transition from the subsonic propeller to the direct accretion regime is  
\begin{equation}
P_{\rm s3}\gtrsim4.5 \dot{M}_{-6}^{-15/21} a_{\rm 10d}^{30/21} v_{8}^{60/21} \mu_{33}^{16/21}
\simeq5.5\dot{M}_{15}^{-15/21}\mu_{33}^{16/21}. 
\label{eq:pspinlimsub}
\end{equation} 

In Fig.~\ref{fig:trans} the above equations are represented as lines separating different
regimes. In panels (a) and (b) we fixed $\mu_{33}$ and v$_{8}$ and investigated 
the different regimes in the  P$_{\rm s3}$-$\dot{M}_{-6}$ plane. In panels (c), (d), (e) and (f), instead, 
P$_{\rm s3}$ and $\mu_{33}$ were fixed and the relevant regimes shown in the 
v$_{8}$-$\dot{M}_{-6}$ plane. 
Below we summarise the different regimes that a system attains in response to
variations of $\dot{M}_{\rm w}$, in the different panels.

Panel (a) shows that, for a wind velocity of v$_{8}$=1.2, 
a strongly magnetized NS ($\mu_{33}$=0.8)
can undergo a transition between the superKeplerian and subKeplerian 
magnetic inhibition regimes in response to changes 
in the mass loss rate only for typical spin periods $\gtrsim$2000~s. 
When the mass loss rate reaches $\dot{M}_{-6}$$\sim$2.4, 
the direct accretion or subsonic propeller regime sets in, depending on whether 
the spin period is longer or shorter than $\sim$3700~s. 
For spin periods $\lesssim$420~s the direct accretion
regime is not attained for the interval of mass loss rates considered in 
Fig.~\ref{fig:trans} and only transitions between the superKeplerian magnetic 
inhibition and supersonic propeller regime are expected.  

For lower magnetic fields ($\mu_{33}$=0.01), panel (b) shows that only the supersonic 
propeller, subsonic propeller, and direct accretion regime can be attained. 
For spin periods in the range $\sim$60-230~s transitions can occur between all these three regimes, 
whereas systems with spin periods longer than $\sim$1300~s and shorter than $\sim$20~s are expected 
to be in the direct accretion and the supersonic propeller regime, respectively. 
For $\mu_{33}$=0.01 transitions to the superKeplerian and subKeplerian magnetic 
inhibition regimes cannot take place because the magnetospheric radius is too small to  
exceed the accretion radius. 

In panel (c) ($\mu_{33}$=0.8 and P$_{\rm s3}$=1), transitions can occur virtually 
between all the regimes described in \S~\ref{sec:model}. 
In particular, for v$_{8}$ in the range 0.9-1.5 transitions are expected to take place between 
the superKeplerian magnetic inhibition, the supersonic and subsonic 
propeller, and the direct accretion regime, as the mass loss rate increases from 
$\dot{M}_{-6}$$\sim$0.1 to $\dot{M}_{-6}$$\sim$100. 
For velocities v$_{8}$$<$0.9 transitions can occur only between the supersonic propeller, 
the subsonic propeller and the direct accretion regime, whereas transitions 
to the superKeplerian and subKeplerian magnetic inhibition regimes  
are impeded by the fact that the accretion radius cannot be overtaken by $R_{\rm M}$.  
On the contrary, for v$_{8}$$>$1.5, the magnetospheric radius is located beyond  
the accretion radius for any considered value of the mass loss rate, and thus transitions can 
take place only between the superKeplerian and subKeplerian magnetic inhibition regimes. 
Similar considerations apply to panels (d), (e), and (f). 
 
For a system with $\mu_{33}$=0.01 and P$_{\rm s3}$=1 (panel (d)), the superKeplerian magnetic 
inhibition regime never occurs, because the magnetic field is too low and 
R$_{\rm M}$$<$R$_{\rm co}$ for any 0.1$<$$\dot{M}_{-6}$$<$100. 
Instead, the subKeplerian magnetic 
inhibition regime can be attained for high wind velocities (v$_{8}$=2), 
because R$_{\rm a}$$\propto$v$_{\rm w}^{-2}$. 

In panel (e), for $\mu_{33}$=0.8 and P$_{\rm s3}$=0.1, the magnetospheric radius is 
larger than the corotation radius for the entire range spanned by $\dot{M}_{-6}$.  
Thus accretion onto the NS does not take place \citep[note that in the region 
below the dashed line accretion can occur even if 
R$_{\rm M}$ $\gtrsim$R$_{\rm co}$,][]{pringle}\footnote{This is because, 
in this region, the mass flow rate is so high that Eq.~\ref{eq:consistencysuper} 
is violated (the supersonic propeller is no longer self-consistent), 
and convective motions are damped by radiative cooling.}.  

Finally, in panel (f) we show the transitions for a system with 
$\mu_{33}$=0.01 and P$_{\rm s3}$=0.1~s. 
In this case all the regimes described in \S~\ref{sec:model} are 
present in the figure, similar to the case of panel (c). 
However, the region corresponding to the subsonic propeller regime is larger,
such that there is only a modest range of (fixed) velocities for which mass loss rate 
variations in our chosen range (0.1$<$$\dot{M}_{-6}$$<$100) 
can cause transitions through all regimes, from the 
superKeplerian magnetic inhibition to the direct accretion regime. 
As we discuss below, this has important consequences for the 
expected luminosity variations. \\ 

\begin{figure}[t!]
\centering
\includegraphics[height=6.5 cm]{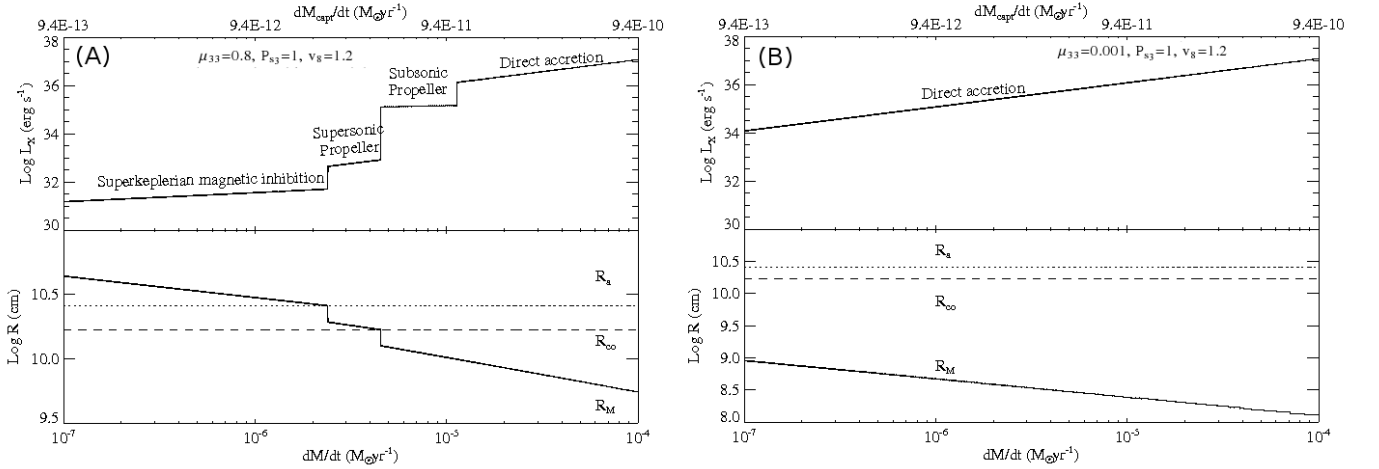}
\smallskip
\caption{(A) \textit{Upper panel}: Variation of the luminosity 
through different regimes, as a function of the mass loss rate from 
the companion star. 
In this case the parameters of the model are fixed at $\mu_{33}$=0.8, 
P$_{\rm s3}$=1, and v$_{8}$=1.2. 
\textit{Lower panel}: Relative position of the magnetospheric radius, 
R$_{\rm M}$ (solid line), with respect to the accretion radius R$_{\rm a}$ 
(dotted line), and the corotation radius R$_{\rm co}$ (dashed line), 
as a function of the mass loss rate from the companion star. \newline
(B) \textit{Upper panel}: Variation of the luminosity 
through different regimes, as a function of the mass loss rate 
from the companion star. 
In this case the parameters of the model are fixed at 
$\mu_{33}$=0.001, P$_{\rm s3}$=1, and v$_{8}$=1.2.  
\textit{Lower panel}: Relative position 
of the magnetospheric radius, R$_{\rm M}$ (solid line), 
with respect to the accretion radius R$_{\rm a}$ (dotted line), 
and corotation radius R$_{\rm co}$ (dashed line), 
as a function of the mass loss rate from the companion star.}  
\label{fig:magn_e_non} 
\end{figure}

We now compute the luminosity swings for some of the 
examples discussed above in a fashion similar 
to what was done in the context of 
centrifugally inhibited accretion in NS  
X-ray transients \citep{corbet96,campana98}.   
Fig.~\ref{fig:magn_e_non}A applies to a system with  
$\mu_{33}$=0.8, P$_{\rm s3}$=1, and v$_{8}$=1.2  
(this corresponds to the case P$_{\rm s3}$=1 of Fig.~\ref{fig:trans} panel (a)). 
The lower panel of this figure shows that, for 0.1$<$$\dot{M}_{-6}$$<$100, 
the magnetospheric radius crosses both the centrifugal (R$_{\rm co}$) and magnetic 
(R$_{\rm a}$) barriers. Correspondingly, the system moves from  
the superKeplerian magnetic inhibition regime, to the supersonic and 
subsonic propeller regime, and, finally, to the direct accretion regime, 
giving rise to a six-decade luminosity swing from $\sim$10$^{31}$ to 
$\sim$10$^{37}$~erg~s$^{-1}$. 
We note that a large part of this swing (about five decades) is attained  
across the transitions from the superKeplerian magnetic inhibition to
the direct accretion regimes, 
which take a mere factor of $\sim$5 variation of $\dot{M}_{\rm w}$. 

In the presence of a standard NS magnetic field ($10^{12}$~G), 
Fig.~\ref{fig:magn_e_non}B  shows that such abrupt luminosity jumps 
are not expected for a very slowly rotating (1000~s) NS 
(the other system parameters are the same 
as those of Fig.~\ref{fig:magn_e_non}A), since the magnetospheric 
radius is smaller than both R$_{\rm a}$ and R$_{\rm co}$, for any 
reasonable value of $\dot{M}_{\rm w}$. Therefore, 
the direct accretion regime applies, with the  
the luminosity proportional to $\dot{M}_{\rm w}$. 

\begin{figure}[t!]
\centering
\includegraphics[height=6.3 cm]{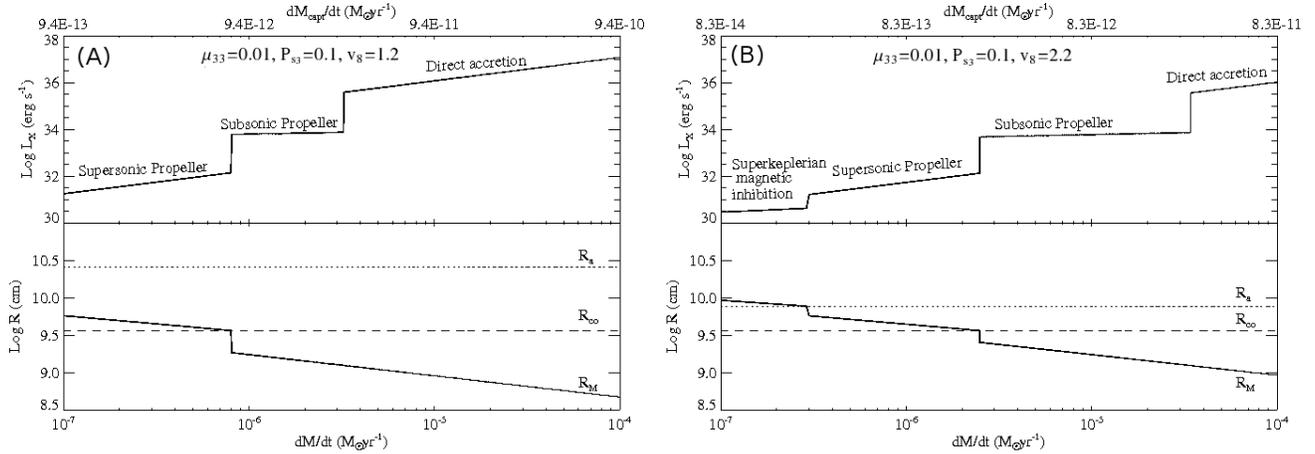}
\smallskip
\caption{(A) \textit{Upper panel}: Variation of the luminosity through 
different regimes, as a function of the mass loss rate. 
In this case the parameters of the model are fixed at 
$\mu_{33}$=0.01, P$_{\rm s3}$=0.1, and v$_{8}$=1.2.  
\textit{Lower panel}: Relative position of the 
magnetospheric radius, R$_{\rm M}$ (solid line), with respect to the 
accretion radius R$_{\rm a}$ (dotted line), and corotation radius 
R$_{\rm co}$ (dashed line), as a function of the mass loss rate from 
the companion star. \newline
(B) Same as (A) but for $\mu_{33}$=0.01, P$_{\rm s3}$=0.1, and 
v$_{8}$=2.2.} 
\label{fig:lshortspin} 
\end{figure}

In Fig.~\ref{fig:lshortspin} we show the 
transitions for a system with $\mu_{33}$=0.01 and P$_{\rm s3}$=0.1.  
The wind velocity is v$_{8}$=1.2 in Fig.~\ref{fig:lshortspin}A, and v$_{8}$=2.2 
in Fig.~\ref{fig:lshortspin}B (see also panel (f) of Fig.~\ref{fig:trans}).  
These two figures show that, for sub-magnetar fields, a 100~s
spinning NS can undergo a transition across the magnetic barrier 
(besides the centrifugal barrier), for suitable parameters 
(a high wind velocity in the case at hand). Such transitions
take place over a more extended interval of mass loss rates. 
For instance Fig.~\ref{fig:lshortspin}B shows that an increase by  
a factor $\sim$100 in the mass loss rate is required, in this case,  
to achieve a factor $\sim$10$^5$ luminosity swing 
comparable with the magnetar case of Fig.~\ref{fig:magn_e_non}A. \\ 

\begin{figure}[t!]
\centering
\includegraphics[height=6.0 cm, width=18.0 cm]{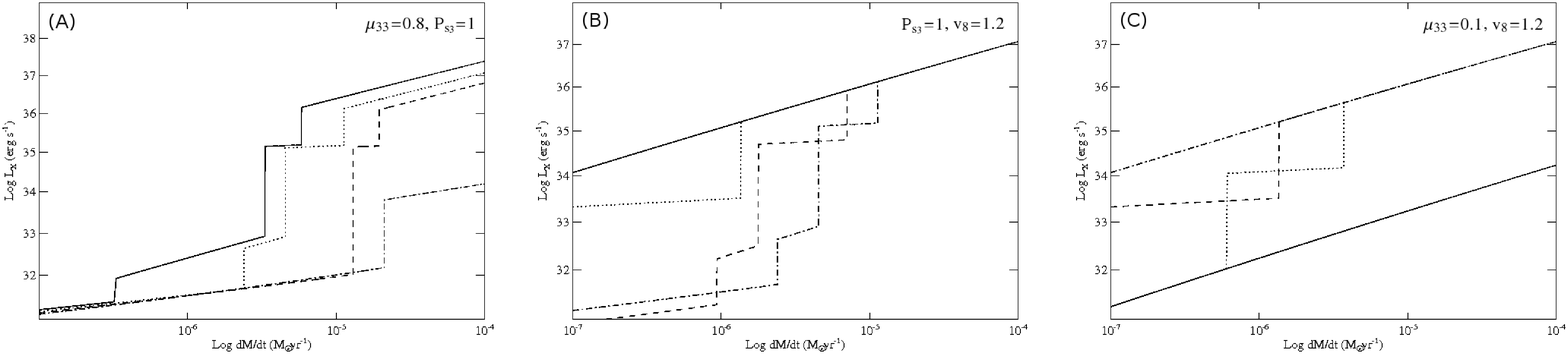}
\smallskip
\caption{Variations of the luminosity through different regimes, 
as a function of the mass loss rate, for different sets of 
parameters. \newline
(A): In this case we fixed $\mu_{33}$=0.8 and P$_{\rm s3}$=1.  
The solid line shows luminosity variations for v$_{8}$=1, the dotted line is for 
v$_{8}$=1.2, and the cases v$_{8}$=1.4 and v$_{8}$=1.8 are represented with 
a dashed and dot-dashed line, respectively. \newline
(B): Here the fixed parameters are P$_{\rm s3}$=1 and v$_{8}$=1.2. 
The different curves correspond to $\mu_{33}$=0.01 (solid line), 
0.1 (dotted line), 0.5 (dashed line), and 0.8 (dot-dashed line). \newline
(C): In this case we fixed $\mu_{33}$=0.1 and v$_{8}$=1.2. Luminosity variations across 
different regimes are shown for P$_{\rm s3}$=0.01 (solid line), 0.5 (dotted line), 
1 (dashed line), and 6 (dot-dashed line).} 
\label{fig:vwind} 
\end{figure}

In order to illustrate further the role of the magnetic field, spin
period and wind velocity, 
we show in Fig.~\ref{fig:vwind} the way in which the transitions across 
regimes take place, by holding two of the above variables
fixed and stepping the third variable. 
Figure~\ref{fig:vwind}A shows the effect of   
increasing the value of v$_{8}$ from 1 to 1.8
(in turn resulting in a decrease of the accretion radius), in a system with 
$\mu_{33}$=0.8 and P$_{\rm s3}$=1.  
 
The behaviour of the luminosity changes mainly because a different set 
of regimes is involved in each case. 
For v$_{8}$=1 (solid line) the system passes through the superKeplerian 
magnetic inhibition regime, the supersonic, and subsonic propeller regime,  
finally reaching the direct accretion regime at $\dot{M}_{-6}$$\simeq$6. 
In the case v$_{8}$=1.2 (dotted line), 
the transition to the supersonic propeller shifts towards higher mass 
loss rates, 
such that superKeplerian magnetic inhibition applies up to $\dot{M}_{-6}$$<$2. 
Further increasing the wind velocity to v$_{8}$=1.4 (dashed line), 
the system first undergoes a transition to the subsonic propeller 
at $\dot{M}_{-6}$$\simeq$10, bypassing the supersonic propeller    
(this is because for $\dot{M}_{-6}$$\simeq$10 the magnetospheric radius 
defined by Eq.~\ref{eq:rmsuper} is smaller than the corotation radius).  
As the mass loss rate increases further, the direct accretion regime sets in  
for $\dot{M}_{-6}$$\simeq$20. 
In the case v$_{8}$=1.8 (dot-dashed line), the corotation radius exceeds the 
accretion radius. The system is thus in the superKeplerian magnetic inhibition 
regime for $\dot{M}_{-6}$$<$20, while for higher mass 
loss rates the subKeplerian magnetic inhibition regime applies and 
the luminosity is dominated by accretion through the KHI. 
A transition to the direct accretion regime is expected 
for mass loss rates $\dot{M}_{-6}$$>$100. 

Figure~\ref{fig:vwind}B shows the transitions across different regimes 
for selected values of the magnetic field, in a system with 
P$_{\rm s3}$=1 and v$_{8}$=1.2. For the lowest magnetic field in this 
figure ($\mu_{33}$=0.01, solid line), the magnetospheric radius is smaller than 
both the accretion and corotation radius for the whole range of mass loss rates 
spanned in the figure, and the system is always in the direct accretion regime. 
In the case $\mu_{33}$=0.1 (dotted line), the system is in the subsonic propeller for 
$\dot{M}_{-6}$$<$1, while for higher mass inflow rates the direct accretion regime applies. 
For $\mu_{33}$=0.5 (dashed line) the luminosity behaviour becomes more complex, as 
the system goes through the superKeplerian 
magnetic inhibition regime, the supersonic and subsonic propeller regime, and 
eventually reaches the direct accretion regime at $\dot{M}_{-6}$$\simeq$8. 
For $\mu_{33}$=0.8 the same sequence of transitions applies, with 
the entire luminosity swing taking place over a smaller interval 
of mass loss rates.

Finally, in Fig.~\ref{fig:vwind}C we show the effects of increasing 
the spin period in a system 
with $\mu_{33}$=0.1 and v$_{8}$=1.2. For the lowest spin period  
considered here (P$_{\rm s3}$=0.01, solid line), the system 
remains in the supersonic propeller regime 
for the whole range spanned by $\dot{M}_{-6}$. In the case  
$P_{\rm s3}$=0.5 (dotted line) transitions occur from the 
supersonic propeller regime ($\dot{M}_{-6}$$<$0.7), to the subsonic 
propeller (0.7$<$$\dot{M}_{-6}$$<$4) and then to the direct accretion regime 
($\dot{M}_{-6}$$>$4). By further increasing $P_{\rm s3}$ 
to a value of 1 (dashed line), the system goes through the subsonic 
propeller and the 
direct accretion regime, while the supersonic propeller regime does not 
occur due to the longer spin period as compared to the previous case. 
For $P_{\rm s3}$=6 (dot-dashed line) the spin period is so 
high that the direct accretion regime applies for any reasonable value of $\dot{M}_{-6}$. 

The above results show that over a range of values of the 
key parameters $\mu_{33}$, P$_{\rm s3}$, and v$_{8}$,  
large luminosity swings can be achieved with comparatively modest 
changes in the mass loss rate, as the neutron star undergoes 
transitions from one regime to another. 
More generally, these transitions result  
from changes in the relative position
of the accretion, magnetospheric, and corotation radii,
reflecting short term variations of the wind 
velocity and mass loss rate, the only parameters 
that can vary on shorter timescales than secular. 
Therefore, transitions between different regimes take place once the source 
parameters are such that the NS straddles the centrifugal barrier (i.e.  
R$_{\rm M}$$\simeq$R$_{\rm co}$, when R$_{\rm a}$$>$R$_{\rm co}$) 
or the magnetic barrier (i.e. R$_{\rm M}$$\simeq$R$_{\rm a}$, 
when R$_{\rm a}$$<$R$_{\rm co}$) or both 
(i.e. R$_{\rm M}$$\simeq$R$_{\rm co}$$\simeq$R$_{\rm a}$). 
The centrifugal barrier applies to relatively short spin periods.  
It is well known that the longer the spin period of 
a transient neutron star, the higher its magnetic field must be
for the centrifugal barrier to operate 
(see Eqs.~\ref{eq:pspinlim} and~\ref{eq:pspinlimsuper}).
In particular, for periods of hundreds seconds, or longer, magnetic field 
strengths of $\gtrsim$10$^{13}$-10$^{14}$~G are required\footnote{This point 
was already noted in \citet{stella86}.}. 
On the other hand, we have shown that 
neutron stars with even longer spin periods 
and magnetar-like fields are expected to undergo transitions across 
the magnetic barrier and thus are expected to have an 
inherently different ``switch off'' mechanism than short spin period 
systems. A necessary condition for this is that R$_{\rm co}$$>$R$_{\rm a}$, 
which translates into P$_{\rm s3}$$\gtrsim$3v$_{8}^{-3}$. 
The magnetic and centrifugal barriers set in (nearly) simultaneously 
(i.e. R$_{\rm a}$$\simeq$R$_{\rm co}$$\simeq$R$_{\rm M}$) 
for $\mu_{33}$$\simeq$0.3P$_{\rm s3}^{7/6}$$\dot{M}_{15}^{1/2}$.  

Taking into account of all the 
examples discussed in this section, we conclude that: 
\begin{enumerate}
\item Long spin period systems (P$_{\rm s3}$$\gtrsim$1) require magnetar-like 
B-fields ($\mu_{33}$$\gtrsim$0.1) in order for a large luminosity swing   
($\sim$10$^5$) to arise from modest variations in the wind parameters 
(e.g. a factor $\sim$5 in $\dot{M}_{-6}$). These luminosity swings might  
result from transition across different regimes through both the centrifugal 
and magnetic barriers. 
\item Shorter spin period systems (P$_{\rm s3}$$\ll$1) must posses lower magnetic 
fields ($\mu_{33}$$\ll$0.1) for similar transitions to take place.  
Somewhat larger variations in the wind parameters are required 
in order to achieve similar luminosity swings to those of the long period case, and 
transitions between different regimes occur in most cases through the centrifugal barrier. 
\item Few or no transitions are expected for systems with either high magnetic 
fields and short spin periods, or systems with lower magnetic fields and long spin periods. 
In the first case the centrifugal barrier halts the 
inflowing matter at R$_{\rm M}$ and accretion does not take place; such systems might 
thus be observable only at very low (X-ray) luminosity levels 
($\simeq$10$^{32}$-10$^{33}$~erg~s$^{-1}$). 
In the second case R$_{\rm M}$$<$R$_{\rm co}$ for a wide range of wind parameters,  
accretion can take place, and a high persistent luminosity is released  
($\simeq$10$^{35}$-10$^{37}$~erg~s$^{-1}$). 
\end{enumerate}

\section{Application to SFXT sources}
\label{sec:results}
 
In this section we propose that transitions across different regimes 
caused by relatively mild variations of the wind parameters are responsible
for the outbursts of SFXTs. In consideration of the wide range 
of spin periods inferred for SFTXs, the outbursts of these sources are expected
to result from the opening of the magnetic barrier in very long spin period systems and 
the centrifugal barrier in all other systems \citep[see also][]{grebenev07,sidoli07}. 
More crucially, we conclude that slowly spinning SFXTs should host magnetars,
independent of which of the two mechanisms is responsible for their outbursts.

\begin{figure*}[t!]
\centering 
\includegraphics[width=18.0 cm]{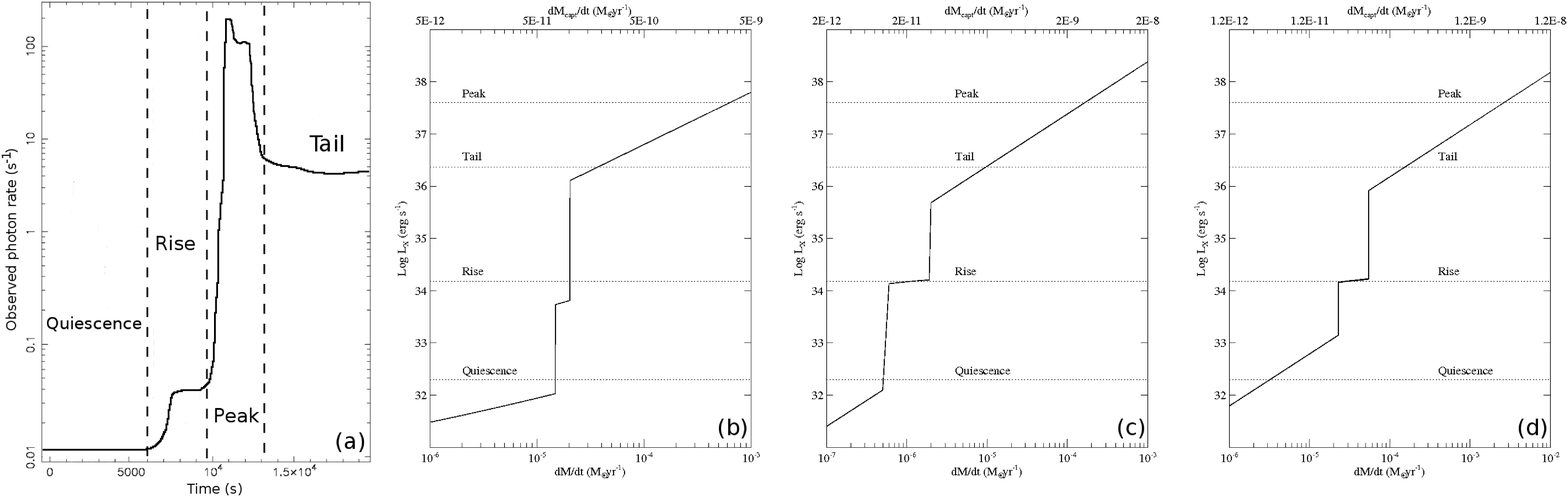}
\smallskip
\caption{Application of the model described in \S~\ref{sec:model} to 
the \IGR\ transition from quiescence to outburst. \newline
{\it Panel (a)}: A schematic representation of the \chan\ light curve 
of \IGR\ obtained by using the segments that were
not affected by pile-up \citep[see both panels of figure 2 in][]{zand05}. 
The different luminosity stages are clearly visible. 
According to \citet{zand05}, the count rates on the y-axis correspond 
to $2\times10^{32}$~erg~s$^{-1}$, 
$1.5\times10^{34}$~erg~s$^{-1}$, $2\times10^{36}$~erg~s$^{-1}$, and 
$4\times10^{37}$~erg~s$^{-1}$, in the quiescence state, the rise state, 
the tail, and the peak of the outburst, respectively. \newline
{\it Panel (b)}: Interpretation of the quiescence to outburst transition of 
\IGR\ in terms of the magnetic barrier model.
The dotted horizontal lines mark the luminosity that 
divide the different regimes. The parameters of the model are fixed at 
$P_{\rm s3}$=1.3, $v_{8}$=1.4, and $\mu_{33}$=1. \newline
{\it Panel (c)}: Interpretation of the quiescence to outburst transition of 
\IGR\ based on the centrifugal barrier model. The parameters of the model are fixed at 
$\mu_{33}$=0.08, $P_{\rm s3}$=0.4, and $v_{8}$=1. \newline
{\it Panel (d)}: Same as panel (c) but here the parameters of the model are fixed at 
$\mu_{33}$=0.001, $P_{\rm s3}$=0.01, and $v_{8}$=2.} 
\label{fig:IGRJ17544} 
\end{figure*}

As a case study we consider \IGR\ \citep{sunyaev03}, a SFXT observed by \chan\ 
during a complex transition to and from a $\sim$1~hour-long outburst,
yielding the first detailed characterization of a SFXT light curve. 
The spin period of \IGR\ is presently unknown. 
\citet{zand05} showed that four different 
stages, with very different luminosity levels, 
could be singled out during the \chan\
observation: (a) a quiescent state with 
L$_{\rm X}$$\simeq$2$\times$10$^{32}$~erg~s$^{-1}$, (b) a rise stage with 
L$_{\rm X}$$\simeq$1.5$\times$10$^{34}$~erg~s$^{-1}$, (c) the outburst peak 
with L$_{\rm X}$$\simeq$4$\times$10$^{37}$~erg~s$^{-1}$, and (d) a post-outburst 
stage (or``tail'') with L$_{\rm X}$$\simeq$2$\times$10$^{36}$~erg~s$^{-1}$ 
\citep[see panel (a) of Fig.~\ref{fig:IGRJ17544}; these luminosities are for a source 
distance of $\sim$3.6 kpc,][]{rahoui08}.   
The maximum luminosity swing observed across these stages was 
a factor of $\gtrsim$6.5$\times$10$^4$. 

Motivated by the evidence for  $>$1000~s periodicities  
in \xte\ and \igg\ (see \S~\ref{sec:properties}), we discuss first 
the possibility that \IGR\ contains a very slowly spinning neutron
star. We use $\mu_{33}$=1, P$_{\rm s3}$=1.3, v$_{8}$=1.4, 
and show in Fig.~\ref{fig:IGRJ17544}(b) the different regimes 
experienced by such a neutron star as a function of the mass loss rate. 
For $\dot{M}_{-6}$$<$20 the above values values give R$_{\rm M}$$>$R$_{\rm a}$ 
and R$_{\rm M}$$>$R$_{\rm co}$, such that superKeplerian magnetic inhibition of accretion applies.  
The expected luminosity 
in this regime, $\sim$10$^{31}$~erg~s$^{-1}$, is likely outshined by the X-ray luminosity 
of the supergiant star \citep[the companion star's luminosity is not shown in 
Fig.~\ref{fig:IGRJ17544}, but it is typically of order $\sim$10$^{32}$~erg~s$^{-1}$,][]
{cassinelli81,berghofer97}. We conclude that the lowest emission state 
(quiescence) of \IGR\ can be explained in this way, with the companion 
star dominating the high energy luminosity \citep{zand05}. 
The rise stage is in good agreement with the subKeplerian magnetic inhibition regime, 
where the luminosity ($\sim$10$^{34}$~erg~s$^{-1}$) is dominated by accretion of matter 
onto the NS due to the KHI. The uncertainty in the value of $h$ translates into 
an upper limit on the luminosity in this regime which is
a factor of $\sim$10 higher than that given above (see \S~3 and Appendix~\ref{app:ht}). 

During the outburst peak the direct accretion regime must apply at a  
mass loss rate of $\dot{M}_{-6}$$=$500. In this interpretation 
direct accretion must also be at work in the outburst tail at 
$\dot{M}_{-6}$$\sim$3, where  
a slight decrease in $\dot{M}_{\rm w}$ would cause the magnetic barrier to close and the 
source to return to quiescence. According to this interpretation, 
if \IGR\ has a spin period of $>$1000~s, then it must host a magnetar. 

Panel (c) of Fig.~\ref{fig:IGRJ17544} shows an alternative interpretation of 
the \IGR\ light curve, where we fixed $\mu_{33}$=0.08, P$_{\rm s3}$=0.4, and 
v$_{8}$=1. 
For this somewhat faster spin (and lower magnetic field), the luminosity variation is mainly 
driven by a transition across the centrifugal barrier (as opposed to the magnetic barrier). 
In this case, the quiescent state corresponds to the supersonic propeller regime 
($\dot{M}_{-6}$ $<$0.6), the rise stage to the subsonic propeller (0.6$<$$\dot{M}_{-6}$$<$2),  
while both the peak of the outburst and the tail take place in the direct accretion regime 
at $\dot{M}_{-6}$=200 and $\dot{M}_{-6}$=10, respectively. 
 
Assuming an even faster NS spin period for \IGR,\ a weaker magnetic field would be
required.   
In panel (d) of Fig.~\ref{fig:IGRJ17544}, we show the results obtained by adopting 
$\mu_{33}$=0.001, P$_{\rm s3}$=0.01, and v$_{8}$=2. The $\sim$10$^{34}$~erg~s$^{-1}$ 
luminosity in the subsonic propeller regime compares well with the luminosity in the 
rise stage. However, the luminosity of the supersonic propeller regime 
is now significantly higher than the quiescence luminosity 
of $\sim$10$^{32}$~erg~s$^{-1}$ (this is consequence of the higher value of 
$\dot{M}_{\rm w}$ for which the supersonic propeller regime is attained in this interpretation). 
We note that, the whole luminosity swing takes place 
for a wider range of mass loss rates, and the outburst peak luminosity requires 
$\dot{M}_{-6}$$\simeq$3000, an extremely high values even for an OB supergiant. 

Interpreting the properties of \IGR\ in terms of a NS 
with a spin periods $\ll$100~s is more difficult.   
For instance, for the subsonic propeller regime to set in, 
the mass loss rate corresponding to the transition across
R$_{\rm M}$=R$_{\rm co}$ must be lower than the limit 
fixed by Eq.~\ref{eq:dotmlimsub}. If instead the transition  
takes place at higher mass loss rate, the system goes directly from the supersonic 
propeller to the direct accretion regime (or vice versa), bypassing the subsonic propeller: 
therefore, the rise stage would remain unexplained. 
Since fast rotating NSs require lower magnetic fields for direct accretion to take place
while in outburst, Eq.~\ref{eq:dotmlimsub} is satisfied 
only for very high wind velocities (v$_{8}$$>$2-3).  
On the other hand, an increase by a factor of $\sim$2 in the wind velocity 
(with respect to the longer spin period solutions) would give a substantially lower 
$\dot{M}_{\rm capt}$, such that the subsonic and the direct accretion regime 
luminosities fall shortwards of the observed values (unless unrealistically high 
mass loss rate are considered).  

Based on the above discussion, we conclude that \IGR\ 
likely hosts a slowly rotating NS, with spin period $>$100~s. 
Whether the magnetic barrier or the centrifugal barrier sets in, causing 
inhibition of accretion away from the outbursts, will depend on whether 
the spin period is longer or shorter then $\sim$1000~s. 
We note that \objectname{IGR J16418-4532}, a $\sim$1240~s pulsating source with a 3.7d orbital 
period, displayed short duration flares similar to those of SFXTs and thus 
might be considered a candidate for hosting an accreting neutron star 
with a magnetar-like field.   
However there is no clear evidence yet that \iigr\ is a transient source, 
since the very low state revealed with \swift\ might well be due to an eclipse
\citep{tomsick04,walter06,corbet06}. 

As another example we discuss the case of \igrjj,\ a SFXTs with  
a spin period of 228~s. The luminosity behaviour  
of this source is still poorly known. An outburst
at 5$\times$10$^{36}$~erg~s$^{-1}$ was observed with \int\  
\citep[assuming a distance of 12.5 kpc,][]{Lutovinov05,smith04}, which did not 
detect the source before the outburst down to a level 
of 5$\times$10$^{35}$~erg~s$^{-1}$. About a week later,  
\xmm\ revealed the source at 5$\times$10$^{34}$~erg~s$^{-1}$  
and discovered the 228~s pulsations \citep{Lutovinov05, zurita04}. 
If the direct accretion regime applied all the way to the 
lowest luminosity level observed so far, then an upper limit 
of $\mu_{33}$=0.004 would be obtained by imposing that 
the neutron star did not enter  
the subsonic propeller regime (see Eq.~\ref{eq:pspinlimsub}). 
On the other hand, if the luminosity measured by 
\xmm\ signalled that the source entered the subsonic propeller
regime, while direct accretion occurred only during the outburst 
detected by \int,\ then a considerably higher 
magnetic field of $\mu_{33}$=0.07 would be required. 

The above discussion emphasizes the importance of determining, 
through extended high sensitivity observations,  
the luminosity at which transitions between different 
source states occur, in particular the lowest luminosity 
level for which direct accretion is still at work. 
In combination with the neutron star spin, this can
be used to infer the neutron star magnetic field.  
Alternatively accretion might take place unimpeded 
at all luminosity levels of SFXTs, a possibility
which requires a very clumpy wind as envisaged in 
other scenarios \citep{negueruela08,zurita07}. In this case 
the neutron star magnetic field can be considerably lower 
than discussed here. 
More extensive studies of these sources (and, by analogy, 
other SFXTs) are clearly required.

\section{Discussion}
\label{sec:discussion}

If the centrifugal barrier operates in \IGR,\ then the activity of this source 
(and by extension that of SFXTs with similar properties) should parallel
that of long spin period X-ray pulsar transients with Be star companions
\citep{stella86}. 
One crucial difference between the two classes,
namely the duration of the outbursts, might well result from the presence in Be systems 
of an accretion disk mediating the flow of matter outside the NS magnetosphere. 
In fact, while in Be systems the star's slow equatorial wind has enough angular
momentum to form such a disk, the supergiant's wind in SFXTs is fast and possesses
only little angular momentum relative to the NS. In the absence of a disk,  
variations in the wind parameters take effect on dynamical timescales, 
whereas in the presence of a disk they are smoothed out  
over viscous timescales. 

If the spin period is sufficiently long in \IGR\ and other SFXTs, the onset of 
the magnetic barrier will inhibit accretion. While steady 
magnetic inhibition of accretion is familiar, e.g. from the earth magnetosphere - 
solar wind interaction, transitions in and out of this regime have not yet been observed, 
to the best of our knowledge. Therefore, very long spin period SFXTs might 
provide the first opportunity to study transitions across such a magnetic barrier. 
Irrespective of whether the centrifugal or magnetic barrier operates in \IGR,\ 
a long spin period would imply a high magnetic field, comparable to 
those inferred for magnetar candidates. 

Scenarios involving magnetars with spin periods well above 1000~s 
have been considered in several studies \citep{ruth, mori, toropina2, liu, zhang04}.   
Moreover a few known accreting X-ray pulsars with unusually long spin 
periods have already been proposed as magnetar candidates 
\citep[see e.g.,][and references therein]{ikhsanov07}. 
However these sources display different properties from SFXTs: some are persistent 
sources; others display week to month-long outbursts; the high spin up measured 
in \igrpatel\ testifies that the accretion flow is likely mediated by a disk \citep{patel07}. 
Therefore some of the features of the model discussed in this paper would 
not be applicable to these sources. 

In the context of wind-fed HMXBs,  
\citet{zhang04} pointed out that magnetars might be hosted in binary systems 
with relatively short orbits ($\sim$1-100~d) 
and long pulse period ($\gtrsim$10$^{3}$-10$^{5}$~s). 
By using evolutionary calculations, these authors 
showed that magnetars would be easily spun-down to such long spin periods by 
the interaction with the wind of the companion star, in less than 10$^{6}$~yr. 
The systems we considered in this paper would   
likely result from a similar evolutionary path. 

Once a system approaches the spin period that is required for 
direct accretion to occur, the short timescale ($\sim$~hours) erratic 
variations that characterize a supergiant's wind  
will cause transitions across different NS regimes.   
Insofar as the average wind properties evolve only secularly, the 
NS spin will then remain locked around such period, alternating spin-up 
intervals during accretion and spin-down intervals during quiescence, 
when accretion is inhibited. 

The relevant spin-up timescale during accretion intervals is 
approximately \citep[see e.g.][]{fkr} 
\begin{equation}
\tau_{\rm su} =  -P_{\rm spin}/\dot{P}_{\rm spin}=\Omega I/(\dot{M}_{\rm capt} l)   
\simeq 8\times10^2 I_{45} v_{8}^{4} P_{\rm 10d} P_{\rm s3}^{-1} \dot{M}_{17}^{-1} ~{\rm yr},     
\label{eq:spinup}
\end{equation}
where l=2$\pi$R$_{\rm a}^2$/(4P$_{\rm orb}$) the specific 
angular momentum of wind matter at the accretion radius, 
I=10$^{45}$I$_{45}$~g~cm$^2$ the NS moment of inertia, and 
$\dot{M}_{17}$=$\dot{M}_{\rm capt}$/10$^{17}$~g~s$^{-1}$ 
(this corresponds to an outburst luminosity 
of $\sim$10$^{37}$erg s$^{-1}$).   

A rough estimate of the spin-down timescale is 
obtained by assuming that most of the quiescence luminosity 
draws from the rotational energy of the NS \citep{pringle}; 
this gives
\begin{equation}  
\tau_{\rm sd}=P_{\rm spin}/\dot{P}_{\rm spin}=I \Omega^2/L_{\rm X}=
1.3\times10^2 I_{45} L_{31}^{-1} P_{\rm s3}^{-2} ~{\rm yr},   
\label{eq:spindown}
\end{equation}
where L$_{31}$ is the luminosity produced by the 
interaction between the magnetosphere and the wind  
in units of 10$^{31}$~erg~s$^{-1}$. 
It is apparent that, for magnetar-like fields and long spin periods 
the above timescales are much shorter than the lifetime of 
the supergiant's strong wind phase. Therefore secular changes 
in the wind parameters and/or the NS magnetic field 
will be easily tracked by the NS spin. 
Moreover since spin-up and spin-down take place on
comparable timescales, it is to be expected that spin-up 
intervals during outbursts are compensated for by 
spin-down during quiescent intervals of 
comparable duration. In other words accretion state of 
long spin period SFXTs would be expected to have 
a duty cycle of order $\sim$0.5 or higher. As we discussed in 
\S~\ref{sec:properties}, evidence for high values of the duty cycle 
is gradually emerging from high sensitivity observations 
of SFXTs.

On the other hand spin-down may also occur during the accretion 
intervals as a results of velocity and density gradients in the 
supergiant's wind that lead to temporary reversals 
of the angular momentum of the captured wind relative to 
the neutron star \citep[note that persistent wind accreting 
X-ray pulsars in HMXBs have long been known to alternate 
spin-up and spin-down intervals, see e.g.][]{henrichs}.
This would tend to favor spin-down, such that the NS spin might 
gradually evolve longwards of the spin period that is required 
to inhibit direct accretion and a transient source becomes a persistent
one. Interestingly, this might apply to \2s,\ a persistent 
X-ray pulsar with a luminosity of $\sim$10$^{36}$~erg~s$^{-1}$,  
which displayed variations by a factor of $\lesssim$10. 
Since its spin period is extremely long ($\sim 10^4$)
$R_{\rm M}$ is smaller than the 
corotation radius and the NS accretes continuously from the relatively 
weak supergiant companion's wind \citep[see also][]{li99}. 

The SFXTs population might thus represent those 
supergiant HMXBs systems that have not (yet) evolved 
away from the spin period at which transitions across 
different NS state can take place. 

We note that for SFXTs to host magnetars their 
dipole magnetic field must retain values in the 
$\mu_{33}$=0.1-1  range for a few 10$^6$~yr, i.e. the typical
timescale from the formation of the NS to the onset 
of the supergiant's strong wind. 
Presently known magnetar candidates (soft gamma repeaters, SGR, 
and anomalous X-ray pulsars, AXP) have estimated ages in the 10$^4$-10$^5$~yr 
range \citep[for a review see e.g.][]{woods}. 
Little is known on the long term evolution of the 
magnetic field of magnetars and different models have 
been proposed which lead to different predictions. 
We note that if the irrotational mode of 
ambipolar diffusion dominates the B-field decay, 
$\mu_{33}$=0.1-0.3 can be expected for ages of a few 10$^6$~yr
\citep{heyl98}.  
 
According to our proposed scenario the combination of a long spin 
period and a very large luminosity swing is indicative 
of the presence of a magnetar. This can be further corroborated 
through other magnetars signatures, such as e.g. proton cyclotron 
features in the X-ray spectrum \citep{zane01} 
or sporadic subsecond bursting activity such 
as that observed in AXPs and SGRs \citep{gavriil02}.  

Finally we remark on the orbital period of SFXTs: in all 
regimes described in \S~\ref{sec:model}, the luminosity 
scales with L$_{\rm X}$$\propto$$P_{\rm orb}^{-4/3}$v$_{\rm w}^{-4}$.  
For our fiducial wind parameters, orbital periods of  
tens of days are required for the transition 
between low and high luminosity states to occur at 
$\simeq$10$^{36}$~erg~s$^{-1}$, a typical  
luminosity for the onset of SFXT outbursts \citep{walter07}. 
This is why throughout this paper we scaled our equations by 
P$_{\rm orb}$=10~d and used the same value in the examples of Figs.~\ref{fig:trans}-\ref{fig:IGRJ17544}. 
\citet{negueruela08} showed that orbital periods around 
$\sim$10~d are compatible with the NSs being embedded in the   
clumpy wind from the supergiant companions, rather 
than in a quasi-continuous wind. In this case SFXT outburst 
durations might be associated with the transit time of a clump
\begin{equation}
\tau_{\rm out}\simeq a/v_{\rm w}=4.2\times10^{4} 
a_{\rm 10d} v_{8}^{-1} ~ {\rm s}, 
\end{equation} 
in reasonable agreement with the observed durations of individual flares 
(see \S~\ref{sec:properties}). 

We conclude that clumpiness of the stellar wind, an often-used 
concept for interpreting SFXT activity, applies to the gating 
scenarios described here as well. The main advantage of introducing 
a gating mechanism rests with the possibility to model 
the very large luminosity swings of 
SFXTs with much milder density (or velocity) 
contrasts in the wind.

\section{Conclusions}
\label{sec:conclusions}

In this paper we reviewed the theory of wind accretion in HMXBs hosting a magnetic
neutron star with a supergiant companion, and considered in some detail the 
interaction processes between the inflowing plasma and the magnetosphere
that are expected to take place when direct accretion onto the 
neutron star surface is inhibited. We then applied this theory to SFXTs   
and showed that their large luminosity swings
between quiescence and outburst (up to a factor of $\sim$10$^{5}$)
can be attained in response to relatively modest variations of the 
wind parameters, provided the system undergoes transitions across different 
regimes. Expanding on earlier work, we found that such transitions can be 
driven mainly by 
either: (a) a centrifugal barrier mechanism, which halts direct accretion 
when the neutron star rotation becomes superKeplerian at the magnetospheric 
radius, a mechanism that has been discussed extensively in Be star X-ray transient 
pulsars, or (b) a magnetic barrier mechanism, when the magnetosphere 
extends beyond the accretion radius. Which mechanism  
and wind interaction regime apply will depend  
sensitively on the NS spin period and magnetic field, 
besides the velocity and mass loss rate in the supergiant's wind. 
In particular, the magnetic barrier mechanism requires 
long spin periods ($\gtrsim$1000~s) coupled with magnetar-like 
fields ($\gtrsim$10$^{14}$~G). 
On the other hand, magnetar-like 
fields would also be required if the centrifugal barrier set in  
at relatively high luminosities ($\gtrsim 10^{36}$~erg~s$^{-1}$)  
in neutron stars with spin periods of hundreds seconds. 

Evidence has been found that the spin periods of a few SFXTs 
might be as long as 1000-2000~s. Motivated by this, we 
presented an interpretation of the 
activity of \IGR\ (whose spin period is unknown) in terms of the   
magnetic barrier by a 1300~s spinning neutron star 
and showed that the luminosity
stages singled out in a \chan\ observation of this source 
are well matched by the different regimes of wind-magnetosphere
interaction expected in this case.
We discussed also an interpretation  
of this source based on the centrifugal barrier and a slightly  shorter 
spin period (400~s), which reproduced the luminosity stages comparably well. 
We emphasise that in both solutions the required magnetic field 
strength ($\gtrsim$10$^{15}$~G and $\gtrsim$8$\times$10$^{13}$~G, respectively) 
are in the magnetar range. 

While the possibility that magnetars are hosted in binary system with supergiant 
companions has been investigated by several authors \citep[e.g.,][]{zhang04,liu},   
clear observational evidence for such extremely high magnetic field neutron
star in binary systems is still missing. According to the present study, 
long spin period SFXTs might provide a new prospective for detecting and studying 
magnetars in binary systems.  

\acknowledgements
EB thanks the CEA Saclay, 
DSM/DAPNIA/Service d'Astrophysique 
for hospitality during part of this work. 
MF acknowledges the French Space
Agency  (CNES) for financial support.  
We would like to thank the referee for useful comments.
This work was partially supported through ASI 
and MUR grants.

\appendix

\section{On the height of the KHI unstable layer}
\label{app:ht}
In this section we expand on the approximation 
h$_{\rm t}$$\sim$R$_{\rm M}$, introduced previously in \S~\ref{sec:intermediate}. 
As discussed by \citet{bur}, the height h$_{\rm t}$ of the layer where matter and magnetic field 
coexist due to the KHI, is mostly determined by the largest wavelength unstable 
mode of the KHI itself. These authors suggested that 
\begin{equation}
h_{\rm t}\simeq h H_{\rm s}, 
\label{eq:ht}
\end{equation} 
where $h$ is a factor of order $\sim$1, and 
H$_{\rm s}$=R$_{\rm M}^2$ k$_{\rm b}$ T(R$_{\rm M}$)/(GM m$_{\rm p}$)  
is the scale height of the 
magnetosheath (note that H$_{\rm s}$ is roughly of the same order as
R$_{\rm M}$$\simeq$10$^{10}$~cm for T$\sim$10$^{8}$~K). 
In the subsonic propeller regime, Eq.~\ref{eq:ht} might be a reasonable assumption 
due to the presence of an extended atmosphere around the magnetospheric boundary; 
on the contrary, it cannot be used in the subKeplerian magnetic 
inhibition regime, where matter flowing toward the NS is shocked very close to the 
magnetospheric boundary. Despite all these uncertainties, we show below that 
the assumption h$_{\rm t}$=h R$_{\rm M}$, with h$\sim$1, gives a conservative 
estimate of the mass accretion rate due to the KHI. 
Using Eq.~\ref{eq:masscons} and the definition 
v$_{\rm conv}$=$\eta_{\rm KH}$v$_{\rm sh}$($\rho_{\rm i}$/$\rho_{\rm e}$)$^{1/2}$
(1+$\rho_{\rm i}$/$\rho_{\rm e}$)$^{-1}$, we derive the equations that define 
the ratio of the densities inside ($\rho_{\rm i}$) and outside ($\rho_{\rm e}$) 
the magnetosphere. These are 
\begin{equation}
(\rho_{\rm i}/\rho_{\rm e})^{1/2}(1+\rho_{\rm i}/\rho_{\rm e})=
0.3 \eta_{\rm KH} h^{-1} R_{\rm M10}^{3/2} P_{\rm s3}^{-1}, 
\label{eq:rho1}
\end{equation} 
if v$_{\rm sh}$=v$_{\rm rot}$, and 
\begin{equation}
(\rho_{\rm i}/\rho_{\rm e})^{1/2}(1+\rho_{\rm i}/\rho_{\rm e})=
0.1 \eta_{\rm KH} h^{-1} R_{\rm M10}^{1/2} v_{8}, 
\label{eq:rho2}
\end{equation} 
if v$_{\rm sh}$=v$_{\rm ps}$.  
Equations~\ref{eq:rho1} and \ref{eq:rho2} show that $\rho_{\rm i}/\rho_{\rm e}$ is an   
increasing function of h$^{-1}$ (for fixed values of v$_{8}$, P$_{\rm s3}$ and R$_{\rm M10}$). 
Therefore, being 
v$_{\rm conv}$$\propto$($\rho_{\rm i}$/$\rho_{\rm e}$)$^{1/2}$(1+$\rho_{\rm i}$/$\rho_{\rm e}$)$^{-1}$ 
and $\dot{M}_{\rm KH}$$\propto$v$_{\rm conv}$, 
the KHI rate of plasma entry inside the magnetosphere is 
also an increasing function of  h$^{-1}$ (provided that $\rho_{\rm i}$$\lesssim$$\rho_{\rm e}$). 
Since the KHI unstable layer is located inside the NS magnetosphere, the maximum height attainable is 
h$_{\rm t}$=R$_{\rm M}$, and thus the approximation used in \S~\ref{sec:intermediate} and 
\S~\ref{sec:subsonic} gives a  lower limit on the mass flow rate controlled by the KHI in both 
the subKeplerian magnetic inhibition and the subsonic propeller regime. 

Note also that the above lower limit does not violate the stability condition   of the 
quasi-static atmosphere in the subsonic propeller regime. 
In fact, following \citet{ikhsanov01b}, this atmosphere can be considered quasi-static  
if the relaxation time scale of the envelope
is less than the drift time scale of the mass flow crossing the magnetospheric boundary.  
In our case matter penetration inside the NS magnetosphere in the subsonic propeller 
regime is mostly provided by the KHI, and the above condition translates into 
$\dot{M}_{\rm KH}$$\lesssim$(R$_{\rm M}$/R$_{\rm co}$)$^{3/2}$$\dot{M}_{\rm capt}$, 
which is satisfied for a wide range of parameters (see Figs.~\ref{fig:magn_e_non}, 
\ref{fig:lshortspin}, and \ref{fig:vwind}).  

An upper limit to the KHI mass flow rate can be obtained by assuming 
$\rho_{\rm i}$/$\rho_{\rm e}$=1 \citep[a solution 
adopted, for example, by][]{ruth}. 
According to \citet{bur}, the density $\rho_{\rm i}$ can be increased 
until the thermal pressure inside the magnetosphere p$_{\rm i}$$\propto$$\rho_{\rm i}$ 
is comparable to the magnetic pressure p$_{\rm m}$=B$^2$(R$_{\rm M}$)/(8$\pi$). When this 
limit is reached, an instability occurs on the lower surface of the unstable 
layer that increases h$_{\rm t}$ until p$_{\rm i}$$<$p$_{\rm m}$ is restored. 
Using the post-shocked gas temperature at R$_{\rm M}$ (see \S~\ref{sec:model}), 
it is shown that $\rho_{\rm i}$/$\rho_{\rm e}$$\sim$1 does not violate the condition 
p$_{\rm i}$$<$p$_{\rm m}$ (at least in the subKeplerian magnetic inhibition regime). 
Therefore, even though we restricted ourself to the lower limit h=1, 
the upper limit on $\dot{M}_{\rm KH}$ might be attainable in some instances.  
The KHI would then provide matter penetration inside the magnetosphere at a rate 
$\sim$$\dot{M}_{\rm capt}$, thus allowing almost all the captured matter to 
accrete onto the NS. A detailed calculation of KHI accretion 
in the subKeplerian magnetic inhibition and subsonic propeller regimes will be reported 
elsewhere.

\section{Radiative losses in the supersonic propeller}
\label{app:supersonic}
As shown by \citet{pringle}, the treatment of the supersonic propeller 
regime (see \S~\ref{sec:supersonic}) is self consistent 
only if the energy input at the base of the atmosphere, due to the 
supersonic rotating NS magnetosphere, is larger than radiative losses 
within the atmosphere itself. 
The range of validity of this assumption can be determined by using the 
convective efficiency parameter
\begin{equation}
\Gamma=\mathfrak{M}_{\rm t}^2 v_{\rm t}(R) t_{\rm br}(R) R^{-1}, 
\end{equation}   
where $\mathfrak{M}_{\rm t}$=v$_{\rm t}$(R)/c$_{\rm s}$(R) is the Mach number, 
v$_{\rm t}$ and c$_{\rm s}$ are the turbulent and sound velocity, and 
t$_{\rm br}$=2$\times$10$^{11}$ T$^{1/2}$m$_{\rm p}$$\rho^{-1}$(R)~s is the bremsstralhung 
cooling time. 
For most of the NS rotational energy dissipated at the magnetospheric 
radius to be convected away through the atmosphere's outer boundary, 
$\Gamma$ should be $\gtrsim$1 across the entire envelope. 
Taking into account that in the supersonic propeller regime  
v$_{\rm t}$$\simeq$c$_{\rm s}$, c$_{\rm s}$$\sim$v$_{\rm ff}$, T$\sim$T$_{\rm ff}$ and $\rho$  
is given by Eq.~\ref{eq:hydro}, one finds $\Gamma$$\propto$R$^{-3/2}$.  
Therefore the said requirement is satisfied when $\Gamma$(R$_{\rm a}$)$\gtrsim$1, i.e.  
\begin{equation}
\dot{M}_{-6}\lesssim 2.2\times10^2 v_{8}^5 a_{\rm 10d}^2.  
\label{eq:consistencysuper}
\end{equation}
For mass loss rates larger than the above limit, radiative losses are not 
negligible and the treatment used in \S~\ref{sec:supersonic} for the 
supersonic propeller regime is no longer self-consistent. 
We checked that the limit of Eq.~\ref{eq:consistencysuper} is never 
exceeded in the cases of interest. We note that a similar value 
was also derived by \citet{ikhsanov02}, but his 
limit is a factor 10 larger than ours. This might be due to the fact that
\citet{ikhsanov02} used $\rho_{\rm w}$ instead of $\rho_{\rm ps}$(1+16/3)$^{1/2}$ 
in the expression for the matter density at R$_{\rm a}$. 

\section{Radiative losses in the subsonic propeller}
\label{app:subsonic}
A similar calculation to that in Appendix~\ref{app:supersonic} in the 
subsonic propeller regime shows that 
$\Gamma$$\propto$R$^{1/2}$, and radiative losses are negligible if    
$\Gamma$(R$_{\rm M}$)$\gtrsim$1, i.e. 
\begin{equation}
\dot{M}_{-6}\lesssim2.8\times10^2 P_{\rm s3}^{-3} 
a_{\rm 10d}^{2} v_8 R_{\rm M10}^{5/2} (1+16 R_{\rm a10}/(5 R_{\rm M10}))^{-3/2}, 
\label{eq:consistencysub}
\end{equation}
that is the same value as that in Eq.~\ref{eq:dotmlimsub} (a somewhat different value was 
obtained by \citet{ikhsanov01a} assuming a density $\simeq$$\rho_{\rm w}$ at R$_{\rm M}$ 
instead of $\rho_{\rm ps}$(1+16R$_{\rm M}$/(5R$_{\rm a}$))$^{3/2}$).

{}

\end{document}